\documentclass[manuscript]{aastex61}

\shorttitle{Homogeneous TTV Analysis of WASP-43 system }
\usepackage{lineno}
\usepackage{amsmath}
\usepackage{graphicx}
\usepackage{xcolor}
\usepackage{longtable}
\usepackage{times}
\usepackage{tablefootnote}
\usepackage{appendix}
\usepackage{lineno}

\begin{document}

\title{INVESTIGATION OF AN ORBITAL DECAY AND GLOBAL MODELING OF THE PLANET WASP-43 \MakeLowercase{b}}

\correspondingauthor{Fatemeh Davoudi; \"Ozg\"ur Ba\c{s}t\"urk}
\email{fdavoudi@znu.ac.ir; obasturk@ankara.edu.tr}

\author{Fatemeh Davoudi}
\affil{Department of Physics, Faculty of Science, University of Zanjan, P. O. Box 45195-313, Zanjan, Iran}

\author{\"Ozg\"ur Ba\c{s}t\"urk}
\affil{Ankara University, Faculty of Science, Astronomy and Space Sciences Department, TR-06100, Tandogan, Ankara, Turkey}

\author{Sel\c{c}uk Yal\c{c}{\i}nkaya}
\affil{Ankara University, Faculty of Science, Astronomy and Space Sciences Department, TR-06100, Tandogan, Ankara, Turkey}

\author{Ekrem M. Esmer}
\affil{Ankara University, Faculty of Science, Astronomy and Space Sciences Department, TR-06100, Tandogan, Ankara, Turkey}

\author{Hossein Safari}
\affil{Department of Physics, Faculty of Science, University of Zanjan, P. O. Box 45195-313, Zanjan, Iran}

\begin{abstract}

WASP-43\,b is one of the most important candidate for detecting an orbital decay. We investigate pieces of evidence for this expectation as variations in its transit timings, based on the ground and space observations. The data set include the transit observations at the T\"UB{\.I}TAK National Observatory of Turkey (TUG) and Transiting Exoplanet Survey Satellite (TESS). We present a global model of the system, based on the most precise photometry from space, ground, and archival radial velocity data. Using the homogenized data set and modeled light curves, we measure the mid-transit times for WASP-43\,b. Our analysis agrees with a linear ephemeris which we refine its light elements for future observations of the system. However, there is a negative difference between the transit timings derived from TESS data in two sectors (09 \& 35) and a hint of an orbital period decrease in the entire data set. Both findings are statistically insignificant due to the short baseline of observations, which prevents us from drawing firm conclusions about the orbital decay of this ultra-short period planet. However, assuming the effect of this decrease of the period in the planet’s orbit, we derive a lower limit for the reduced tidal quality factor as $Q_{\;*}^{'} > (4.01 \pm  1.15) \times 10^5$ from the best-fitting quadratic function. Finally, we calculate a probable rotational period for this system as $7.52$ days from the out-of-transit flux variation in the TESS light curves due to spot modulation.
\end{abstract}


\section{Introduction}\label{sect1}

WASP-43\,b with a period about 0.8 days \citep{Hellier2011} is an example of the ultra-short period (periods of a few days or less) hot-Jupiters \citep{Gillon2014, Hartman2016}. The tidal dissipation of energy between the host star and the planet leads to the orbit shrinking \citep{Levrard2009}. This stellar tidal dissipation is usually stated through the tidal quality factor $Q_{\;*}^{'}$. The orbital energy that must be dissipated to circularize the orbit of an ultra-short period planet can easily exceed the planet’s internal binding energy; this process either makes the planet very bright (because tidal dissipation generates heat) or destroys the orbit of the planet due to orbital decays. Such an  orbital decay is expected from especially those with the smallest orbital sizes comparable to their Roche limits \citep{Ogilvie2014}, which can be detected through transit timing variations (TTVs) \citep{Hellier2009}. $Q_{\;*}^{'}$ can be measured based on the rate of decline in the orbital period over a few years \citep{Birkby2014}.

Studies to date have shown that WASP-43\,b is one of the most promising ultra-short period hot-Jupiters to detect orbital decay in their TTVs, amongst other planets such as WASP-12\,b \citep{Hebb2009}, WASP-4\,b \citep{Wilson2008}, WASP-19\,b \citep{Hebb2010}, WASP-18\,b \citep{Hellier2009}, WST-2\,b \citep{Birkby2014}. So far, WASP-12\,b is the only case with a confirmation of its orbital decay \citep{Yee2020}, and WASP-4\,b and WASP-19\,b are strong candidates \citep{Southworth2019,Patra2020}. These planets are located very close to their host stars orbiting with periods of less than $2.3$ days and orbital distances of less than $0.038$ au.
	
WASP-43\,b was discovered in 2011 on a short-period orbit ($P = 0.813474061(46)$ days) \citep{Wong2020} around its host star, with a semi-major axis of $0.0142 \pm 0.0004$ au \citep{Hellier2011}. This short orbit makes it an excellent candidate for orbital decay, and hence an expectation of a short remaining lifetime of about 8 Myrs if $Q_{\;*}^{'}$ is on the order of $10^{6}$, and 80 Myrs if it is on the order of $10^{7}$ \citep{Hellier2011}. $Q_{\;*}^{'}$ value is a measure of how the host star dissipates energy released by tidal interactions with a close-in planet and therefore depends on its spectral type \citep{Barker2009}. Since WASP-43 is a K7V star \citep{Salz2015} and most of orbital decay candidates orbit host stars with earlier spectral types (F and G), therefore, WASP-43 provides an excellent opportunity to test the theoretical values of $Q_{\;*}^{'}$ in late-type stars in the tidal evolution of ultra-short period hot-Jupiters \citep{Hellier2011}.

The first TTV studies for WASP-43b showed mid-transit times in the TTV diagram to be consistent with a linear ephemeris, and there was no sign of TTVs \citep{Gillon2012, Maciejewski2013}. \cite{Blecic2014} added some mid-transit times from amateur observations to the TTV diagram. Although they found the quadratic model superior to a linear model based on comparison statistics, they computed the lower limit for the reduced tidal quality factor as $Q_{\;*}^{'} > 12000$ using the upper limit on the best fitting quadratic term that is much lower than the canonical $Q_{\;*}^{'}$  for such a system. \cite{Chen2014} added some more mid-transit times and found that a linear model is appropriate and noted that there is no significant TTV in the system. \cite{Murgas2014} saw a decrease in the period in their TTV diagram and claimed that the data can be better fitted by using a quadratic function; from which they computed the rate of period change as $\dot{P}= - 0.15 \pm 0.06 \; s\; yr^{-1}$. They emphasized a need for a more homogeneous analysis of more transit timing data after several years. \cite{Ricci2015} added several more mid-transit times and reported that their ephemeris analysis gives a period consistent with the previous results, rejecting an improvement by a quadratic model. \cite{Jiang2016} worked on 31 previous and eight new mid-transit times through homogeneous analysis and concluded that TTVs exist due to a slow orbital decay. They presented $\dot{P}= - 0.02890795 \pm 0.0077254\; s\; yr^{-1}$ for the orbital decay and found a $Q_{\;*}^{'}$-value on the order of $10^{5}$.

Then, \cite{Hoyer2016} did a comprehensive analysis of 72 mid-transit times and measured $\dot{P}= - 0.02 \pm 6.6 \; ms\; yr^{-1}$ which is almost zero, and hence consistent with a constant period. They claimed that $Q_{\;*}^{'}< 10^{5}$ can be discarded for WASP-43\,b, and after about five years (about 4500 epochs), it would be possible to probe the $Q_{\;*}^{'}< 10^{6}$ limits. \cite{Stevenson2017} stated that the orbital period is increasing ($\dot{P} = 0.009 \pm 0.004\; s \; yr^{-1}$), against all the expectations of an orbital decay for WASP-43\,b. Using their best-quadratic fit to their TTV diagram, \cite{Zhao2018} claimed that there could be a long-term TTV due to orbital decay with a rate of $\dot{P} = - 0.005248 \pm  0.001714 \; s\;yr^{-1}$ that gives a lower limit for the reduced tidal quality factor as  $Q_{\;*}^{'}\ge 1.5\times 10^{5}$. \cite{Patra2020} added three more mid-transit times and found an increase in the orbital period with a rate of $\frac {dP}{dt} = (1.9 \pm 0.6) \times 10^{-10}$ and estimated the lower limit on $Q_{\;*}^{'} > (2.1 \pm 1.4 )\times 10^{5}$ with $95\% $ confidence. Finally, TESS observations in sector-09 were added to the TTV diagram by \cite{Wong2020}. They refined the ephemeris and derived  $\dot{P}$ less than 5.6 miliseconds per year.

After a decade of observations and analyses, the nature of TTVs observed in the WASP-43\,b system is still not understood well, and there are contradictory results. Since the data used in the literature has mostly been insufficient both in number and quality to reach firm conclusions, a careful review of previously analyzed light curves, together with new ground-based observations and super-precise new TESS light curves within a homogeneous measurement and analysis method is needed. Such a complete study based on a homogeneous data set will help us better estimate the value of $\dot{P}$ and the reduced tidal quality factor $Q_{\;*}^{'}$. This can also help estimate the remaining lifetime of WASP-43\,b and understand its orbital evolution more clearly. We also aim at a homogeneous and simultaneous study of the most precise transit light curves, archive radial velocity observations, broadband magnitudes of the system to obtain their global model, and refine system parameters.

This paper is organized as follows. Section \ref{sec:obs} provides the details of our own observations and TESS data. Section \ref{sec:analysis} details the stellar parameters of WASP-43, global modeling, construction of the TTV diagram, and the analysis of the out-of-transit variability in sub-sections. Section \ref{sec:conc} discusses the findings, compares them to the previously found values, and finally puts the results in a perspective of an expectancy of orbital decay from the most important candidates, including WASP-43\,b, amongst other $\sim400$ hot-Jupiters.
 
\section{Observations and Data Reduction}\label{sec:obs}

\subsection{T100 Telescope Data}
We observed this system on 14 April 2015 using a 1-meter telescope (T100) at the TÜBİTAK National Observatory, with an SI-1100 CCD camera ($4096 \times 4096$ pixel array with a pixel length of $15\mu$) that was kept stable at an average temperature of $-95^{\circ} C$ CCD with a cryo-cooler. A Bessel-R filter has been employed during the observation with 60 seconds as exposure time. $2\times2$-binning was applied to reduce the readout time of the images, and the telescope was slightly defocused to increase the signal-to-noise ratio as described in \cite{Basturk2015}. CCD images were reduced, and aperture photometry concerning an ensemble of carefully selected comparison stars according to their positions, magnitudes, and colors was performed in the standard manner by using the AstroImageJ (AIJ) software package \citep{Collins2017}. All the mid-exposure timings were converted to the Dynamical Barycentric Julian Day ($BJD_{TDB}$). We also detrended the light curve for the airmass effect and normalized the fluxes to the out-of-transit level within AIJ. The light curve is presented in Figure-\ref{fig:gm_lcs} which is marked as TUG(R).

\subsection{TESS Data}
 Transiting Exoplanet Survey Satellite (TESS) has observed WASP-43 (TIC 36734222, TESS magnitude 11.118) at a 120-sec cadence using a red-optical bandpass centered on the Cousins$-I_c$ band. Observations have been performed in two-time intervals, sector-09 that started on 28 February 2019 and ended on 26 March 2019 \citep{Wong2020} using camera 1 and CCD 1, and sector-35 using camera 1 and CCD 2 between 9 February 2021 and 7 March 2021.

We downloaded the calibrated TESS data files from Mikulski Space Telescope Archive (MAST). We extracted the TESS light curves detrended by the TESS Science Processing Operations Center (SPOC) pipeline \citep{Jenkins2016} from the files that provide TESS data validation timeseries (dvt files). We used our python codes based on the functions of the lightkurve package \citep{Cardoso2018}. We removed the obvious outliers due to instrumental artifacts and cut the entire data set into smaller chunks centered on individual transits. We present phase folded transit light curves combined from 24 transits of sector-35 and 26 transits of sector-09 in Figure-\ref{fig:gm_lcs}.  
\\

\section{Analysis}\label{sec:analysis}
\subsection{Stellar Parameters}
We fitted the spectral energy distribution (SED) of the star using MESA Isochrones \& Stellar Tracks (MIST) bolometric correction grid\footnote{$http://waps.cfa.harvard.edu/MIST/model\_grids.html$} \citep{choi2016} with {\sc eoxfast}-v2 \citep{eastman2019} based on broadband photometry from ground-based and space-borne observations in different passbands covering the widest wavelength interval possible (listed in Table~\ref{passbands}). We also provided parallax measurements from Gaia Data Release-2 (DR2)  \citep{gaia2016, gaia2018} as a Gaussian prior after the addition of an offset value ($0.082^{\prime\prime}$) noticed by \cite{stas2018}. We preferred Gaia DR2 measurements over Early Data Release-3 (EDR3) Gaia because the offset value on DR2 parallaxes is better known at the time of this study. We also propagated the offset to the uncertainty in parallax as $0.033^{\prime\prime}$. We allowed the V-band extinction value ($A_V$) to vary as a free parameter, but limited it to the line of sight value given by \cite{schlegel1998}. We used atmospheric parameters from the most recent work \citep{sousa2018}, which is dedicated to measuring stellar parameters of exoplanet host stars, as a Gaussian prior to our analysis. However, T$_{\rm eff}$ value from \cite{sousa2018} was significantly higher than previous studies and our SED results. This could be caused due to  reasons described in \cite{tsantaki2013}. \cite{tsantaki2013} found that Stars below 5000 K suffer from blends; hence the T$_{\rm eff}$ determination techniques depending on equivalent widths could lead to inaccurate results. In addition, these stars have very few Fe-II lines to use in an analysis based on ionization balance, so the values derived from this method could also suffer from low number statistics. On the other hand, \cite{tsantaki2013} proved that the Synthetic spectra fitting technique, especially in the infrared region, provided much more accurate results for cold stars. We noticed that synthetic spectra also agree (in terms of SED fitting) with other wavelength regions (i.e., optic region) if the V-band extinction is small. The point of SED fitting is to get T$_{\rm eff}$ and radius of the star with less dependence on theoretical models with the help of the super-precise Gaia parallax. We intended to use only radius value as a Gaussian prior in the global modeling, but due to the wide range of value reported for T$_{\rm eff}$ in previous studies and the perfect agreement of the model in the infrared and other regions in our SED analysis, we also used our T$_{\rm eff}$ value from the SED fitting for global modeling. Error bars for both T$_{\rm eff}$ and R$_{\star}$ values are calculated according to \cite{tayar2020}. Our results for T$_{\rm eff}$ from the SED analysis are listed in Table-\ref{teff_comparison} in comparison with previous studies.

\begin{table*}[ht]
\centering
\caption{Effective Temperature Value for WASP-43}    
\centering
\resizebox{\textwidth}{!}{\begin{tabular}{ccccccc}
 \hline
 \hline
		& This work &\cite{sousa2018} &\citep{Esposito2017}&\cite{Chen2014}& \cite{Gillon2012}&\cite{Hellier2011} \\
		\hline
		T$_\text{eff}$&$4195\pm84$&$4798\pm216$&$4500\pm100$ &$4536\pm98$&$4520\pm120$&$4400\pm200$\\
		Method& SED&EW of Iron Lines&EW of Iron Lines&
		G12\footnote{Parameter taken from \cite{Gillon2012} and used as a prior used in stellar evolution code} + Stellar Evolution&H11\footnote{Parameter taken from \cite{Hellier2011} and used as a prior in stellar evolution code} + Stellar Evolution&H$_{\alpha}$ fitting\\
        \hline
\end{tabular}}
\label{teff_comparison}
\end{table*}

\subsection{Global Modeling}
We modelled every light curve, we have selected according to the criteria we detail in Section-\ref{subsection:ttvdata} with {\sc exofast-v1} \citep{Eastman2013} to get the mid-transit times, calculate the photon noise rate (PNR, \citealp{fulton2011}) and red noise parameter $\beta$ \citep{winn2008} as described in \cite{basturk2020}. Then we selected the "optimum" light curves in each passband with the lowest PNR value acquired by a single telescope, the $\beta$ and $\chi^2_{\nu}$ values of which we then compared with each other to pick the "best" light curve among them in each of the passbands. We phase-folded the TESS transit light curves in each sector using the periods provided by the SPOC pipeline in the dvt files, which we then binned so that we have a single short-cadence transit light curve to increase the precision and reduce the integration time in the global modelling. We picked J, H, K, g', i', r', z' \citep{Chen2014}, I+z \citep{Gillon2012}, V-band \citep{Ricci2015}, Spitzer 4.5 $\mu$m \citep{Stevenson2017}, R-band (T100 observation), and finally the binned-TESS light curves for analysis. Modeling light curves in different passbands increased the range in depth contrasts while the binned-TESS light curve increased the precision. We also made use of every radial velocity data available \citep{Hellier2011, Gillon2012, Esposito2017} in the literature for our global modeling. We employed the Spitzer phase-curve from \cite{Stevenson2017} that includes transit and two occultation-observations to better constrain the eccentricity of the orbit of WASP-43\,b. We used T$_{\rm eff}$ and R$_{\star}$ values from our SED-model, and the metallicity value provided by \cite{sousa2018} as Gaussian priors during our global modeling. 

We simultaneously fitted the radial velocity and selected light curves in line with the stellar evolution tracks of the host star until the parameters are well mixed with the {\sc exofast-v2} software \citep{eastman2019}. Since the mean density of the host star ($\rho_*$) can be better constrained from transits to derive the mass of the star, we did not use $log~{\rm g}$ value which has only a subtle effect on stellar spectra from which they are derived, in the stellar evolution model. We obtained two global models, in one which we enforced a circular orbit assumption while we adjusted the eccentricity in the other. Bayesian Information Criteria (BIC) favored the circular orbit with $\Delta$BIC = 2.13. However, the small $\Delta$BIC value shows that the success of both models to represent the data is similar. WASP-43\,b's orbit could have a very small eccentricity, but more precise RV and photometric-occultation data are required to measure it. Therefore, we selected to adopt the simplistic (circular) model as a solution over the eccentric model, for which we do not have clear evidence, despite many RV data points and Spitzer-occultation observations.

\begin{figure}[ht]
\begin{center}
\includegraphics{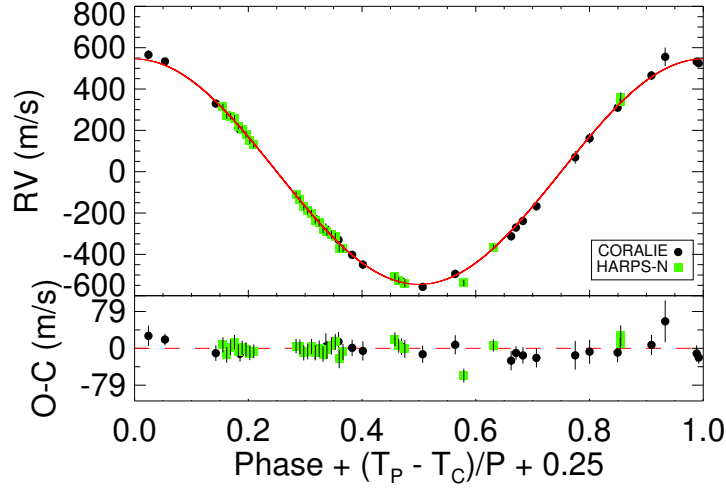}
\end{center}
\caption{Radial velocity observations from CORALIE (black dots) and HARPS-N (green squares) with their error bars with respect to the orbital phase. Red continuous curve is the best-fitting Keplerian model. Residuals are given on the bottom panel.}
\label{fig:rv}
\end{figure}

\begin{figure}[ht]
\includegraphics[width=\columnwidth]{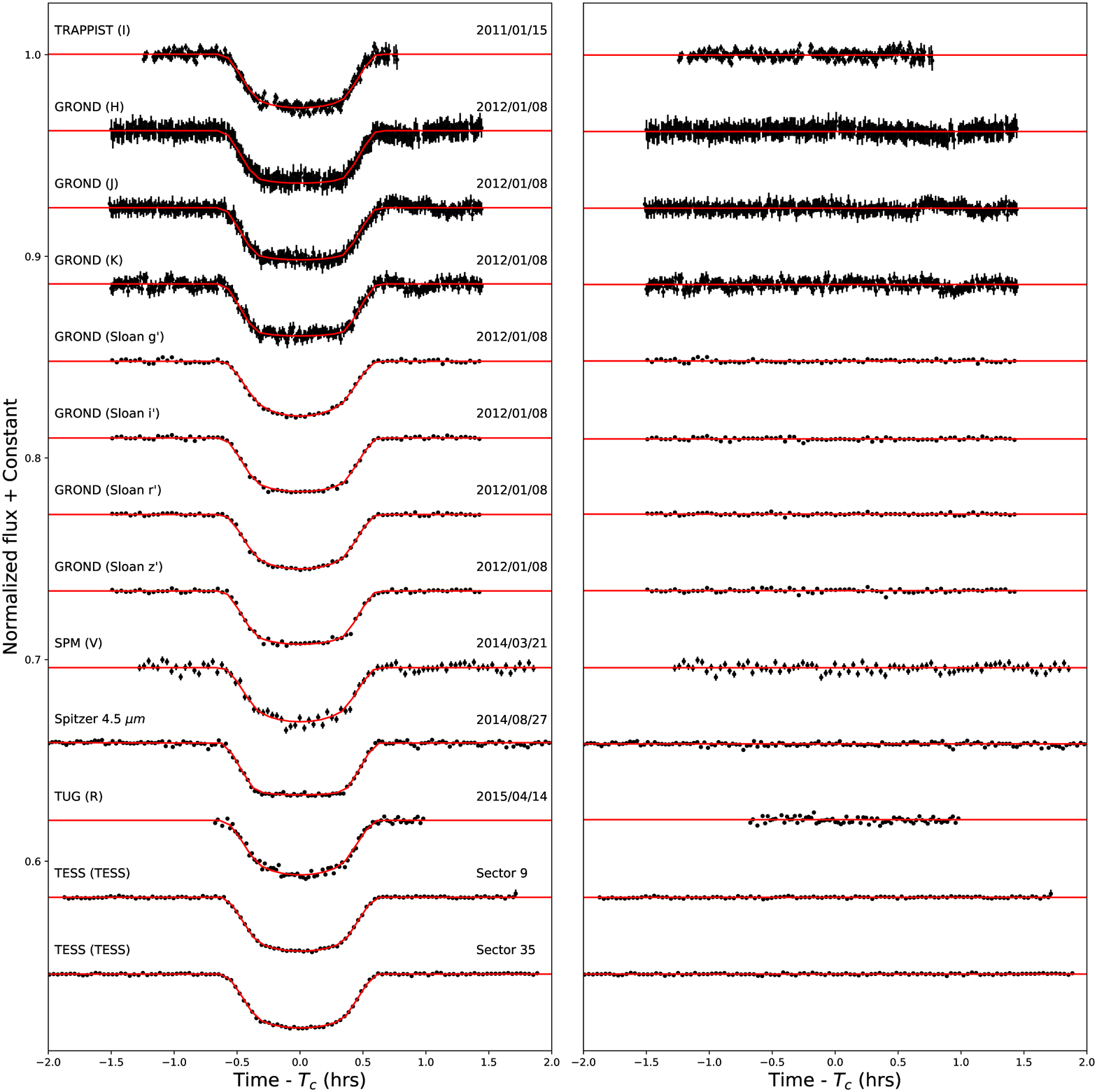}
    \caption{Transit light curves of WASP-43 b used in the global modelling. Observational data are provided on the left panel with the black error bars, red continuous curves are their individual {\sc exofast-v2} models. Residual from them are provided on the right panel. The TESS light curves are phase-folded.}
    \label{fig:gm_lcs}
\end{figure}

Radial velocity and light curves used in the global modelling are provided together with their models in Figures-\ref{fig:rv} \& \ref{fig:gm_lcs}, respectively. Our results from the global modelling are presented in Table-\ref{global_parameters}. Limb darkening parameters are provided within Table-\ref{tab:wave} in the Appendix section.

\begin{table}[ht]
\centering
\footnotesize
\caption{Passband Brightnesses of WASP-43.}
\label{tab2}
\begin{tabular}{ccc}  
\hline
\hline
Passband & $\lambda_{eff}(\AA)$ & Magnitude \\
\hline
\multicolumn{3}{l}{APASS-DR9 \citep{henden2016}}\\
\hline
Johnson B & 4360.0 & $13.796\pm0.022$ \\
Johnson V & 5445.8 & $12.464\pm0.028$ \\
SDSS g' & 4640.4 & $13.182\pm0.061$ \\
SDSS r' & 6122.3 & $11.898\pm0.043$ \\
SDSS i' & 7439.5 & $11.408\pm0.035$ \\
\hline
\multicolumn{3}{l}{GALEX \citep{bianchi2017}}\\
\hline
galNUV & 2274.4 & $21.3355\pm0.2166$ \\
\hline
\multicolumn{3}{l}{2MASS \citep{cutri2013}}\\
\hline
$J_{2MASS}$ & 12350.0 & $9.995\pm0.024$ \\
$H_{2MASS}$ & 16620.0 & $9.397\pm0.025$ \\
$K_{2MASS}$ & 21590.0 & $9.267\pm0.026$ \\
\hline
\multicolumn{3}{l}{All WISE \citep{cutri2013}}\\
\hline
WISE1 & 33526.0 & $9.152\pm0.023$ \\
WISE2 & 46028.0 & $9.225\pm0.020$ \\
WISE3 & 115608.0 & $9.134\pm0.032$ \\
\hline
\multicolumn{3}{l}{Pan-STARRS1 \citep{chambers2016}}\\
\hline
gPS & 4775.6 & $12.883\pm0.061$ \\
rPS & 6129.5 & $11.968\pm0.061$ \\
iPS & 7484.6 & $11.543\pm0.061$ \\
zPS & 8657.8 & $11.279\pm0.061$ \\
yPS & 9603.1 & $11.142\pm0.061$ \\
\hline
\label{passbands}
\end{tabular}
\end{table}

\providecommand{\bjdtdb}{\ensuremath{\rm {BJD_{TDB}}}}
\providecommand{\feh}{\ensuremath{\left[{\rm Fe}/{\rm H}\right]}}
\providecommand{\teff}{\ensuremath{T_{\rm eff}}}
\providecommand{\teq}{\ensuremath{T_{\rm eq}}}
\providecommand{\ecosw}{\ensuremath{e\cos{\omega_*}}}
\providecommand{\esinw}{\ensuremath{e\sin{\omega_*}}}
\providecommand{\msun}{\ensuremath{\,M_\Sun}}
\providecommand{\rsun}{\ensuremath{\,R_\Sun}}
\providecommand{\lsun}{\ensuremath{\,L_\Sun}}
\providecommand{\mj}{\ensuremath{\,M_{\rm J}}}
\providecommand{\rj}{\ensuremath{\,R_{\rm J}}}
\providecommand{\me}{\ensuremath{\,M_{\rm E}}}
\providecommand{\re}{\ensuremath{\,R_{\rm E}}}
\providecommand{\fave}{\langle F \rangle}
\providecommand{\fluxcgs}{10$^9$ erg s$^{-1}$ cm$^{-2}$}

\begin{table*}
\small
\centering
\caption{Median values and 1-standard deviations for the parameters of the star WASP-43 system.}
\setlength{\tabcolsep}{12pt}
\renewcommand{\arraystretch}{0.8}
\begin{tabular}{llc}
\hline \hline
Symbol & Parameter (Unit) & Value\\
\hline
\multicolumn{3}{l}{Stellar Parameters:} \\
\hline
$M_*$&Mass (\msun)&$0.646^{+0.026}_{-0.025}$\\
$R_*$&Radius (\rsun)&$0.6537^{+0.0089}_{-0.0089}$\\
$L_*$&Luminosity (\lsun)&$0.1096^{+0.0079}_{-0.0073}$\\
$\rho_*$&Density (cgs)&$3.26^{+0.052}_{-0.05}$\\
$\log{g}$&Surface gravity (cgs)&$4.6174^{+0.0077}_{-0.0075}$\\
$T_{\rm eff}$&Effective Temperature (K)&$4108^{+65}_{-64}$\\
$[{\rm Fe/H}]$&Metallicity (dex)&$-0.111^{+0.071}_{-0.073}$\\
$[{\rm Fe/H}]_{0}$&Initial Metallicity &$-0.086^{+0.074}_{-0.078}$\\
$Age_{iso}$&Isochrone Age (Gyr)&$8.8^{+3.5}_{-4.8}$\\
$Age_{gyro}$&Gyrochronology Age (Gyr)&$0.216\pm0.004$\\
$EEP$&Equal Evolutionary Point &$330.5^{+8.6}_{-21}$\\
$A_V$&V-band extinction (mag)&$0.082^{+0.037}_{-0.050}$\\
$\varpi$&Parallax (mas)&$11.580^{+0.075}_{-0.079}$\\
$d$&Distance (pc)&$86.36^{+0.59}_{-0.55}$\\
\hline
\multicolumn{3}{l}{Planetary Parameters:}\\
\hline
$P$&Period (days)&$0.813474075\pm0.000000012$\\
$R_{\rm p}$&Radius (\rj)&$1.005\pm0.014$\\
$M_{\rm p}$&Mass (\mj)&$1.907\pm0.052$\\
$a$&Semi-major axis (au)&$0.01476^{+0.00019}_{-0.00019}$\\
$i$&Inclination (Degrees)&$82.18^{+0.11}_{-0.11}$\\
$T_{eq}$&Equilibrium temperature (K)&$1318\pm20$\\
$K$&RV semi-amplitude (m/s)&$549.5\pm3.8$\\
$\rho_{\rm p}$&Density (cgs)&$2.327^{+0.058}_{-0.056}$\\
$logg_{\rm p}$&Surface gravity &$3.6699^{+0.0067}_{-0.0066}$\\
$\Theta$&Safronov Number &$0.0866^{+0.0014}_{-0.0014}$\\
$\fave$&Incident Flux (\fluxcgs)&$0.686^{+0.042}_{-0.040}$\\
\hline
\multicolumn{3}{l}{Transit Parameters:}\\
\hline
$b$&Transit impact parameter &$0.6604^{+0.0057}_{-0.0059}$\\
$R_P/R_*$&Radius of planet in stellar radii&$0.15805^{+0.00034}_{-0.00035}$\\
$a/R_*$&Semi-major axis in stellar radii &$4.854^{+0.026}_{-0.025}$\\
$\tau$&Ingress/egress transit duration (days)&$0.01169\pm0.00017$\\
$T_{14}$&Total transit duration (days)&$0.05157\pm0.00013$\\
\hline
\multicolumn{3}{l}{Occultation Parameters:}\\
\hline
$A_T$&Thermal emission from the planet (ppm)&$2020\pm110$\\
$A_R$&Reflection from the planet (ppm)&$1554^{+42}_{-43}$\\
$\delta_{S}$&Measured eclipse depth (ppm)&$3580\pm120$\\
\hline
\label{global_parameters}
\end{tabular}
\end{table*}

\subsection{Homogenization of the TTV data}
\label{subsection:ttvdata}
We collected a complete set of light curves acquired from the ground and space for this system from the literature, open databases providing observations of amateur \& professional astronomers, such as Exoplanet Transit Database (ETD)\footnote{http://var2.astro.cz} and the MAST portal. We have searched for appropriate observations in the ExoClock\footnote{https://www.exoclock.space/database/planets/} database as well but were not able to find a quality light curve assuming our selection criteria among those not listed in the ETD. We discarded incomplete or low-quality cases to improve transit timing precision. In addition, we detrended the data for all linear effects and converted the timings to $BJD_{TDB}$. We have developed a Python code based on the {\sc astropy} python package \citep{Astropy2018} for this purpose. 

{\sc exofast-v1} was used to model all the light curves and derive the mid-transit times homogeneously. The passband of each observation was provided for the priors of quadratic limb darkening coefficients together with the normalized light curve of the system to out-of-transit flux in the close-vicinity of the ingress and egress. Priors for the stellar parameters and some of the critical transit parameters (orbital period, inclination, and eccentricity) were taken from \cite{Gillon2012, Hellier2011} for the system through their web interface hosting {\sc exofast-v1}. The modeling software then followed a least-squares approach to provide the model parameters. 

We carefully examined all output parameters, the depth, and the duration of the transit in particular for the sanity of the model. The presence of red and white noise leads to inaccurate estimations of the transit length and depth and inaccuracies, in the calculations of the mid-transit times. We compared the derived depth and duration of transits derived from our models to their predicted values reported in \citep{Hellier2011} in the range of $3\sigma$. In cases where this discrepancy was due to significant noise in the data, the light curve was discarded. The Hubble and Spitzer mid-transit times were taken directly from the literature \citep{Stevenson2014, Stevenson2017}. Derived mid-transit times and their uncertainties were used to form the TTV diagram by computing the differences between them and their expected values with reference to a selected mid-transit time \citep{Weaver2020, Wong2020}. Then we supplemented our data with mid-eclipse times derived from high-quality occultation observations, including 5 Hubble and 8 Spitzer observations which are presented with red triangles on the TTV diagram (section-\ref{ttvAnalysis}). We obtained these mid-occultation timings directly from the literature \citep{Stevenson2014, Stevenson2017}. The details of described TTV diagram, including mid-transit and mid-occultation times, TTV values, and some transit parameters derived from {\sc exofast-v1} are presented in Table-\ref{ttvanalysis} in the Appendix section. Three mid-occultation times were not included in the analysis of the TTV diagram, because they were outliers, and we could not find the reason for this discrepancy. They are marked with an asterisk symbol ($*$) in the electronic table. Two of them are indicated in gray as ignored occultations on the TTV diagram (Figure-\ref{fig:ttv}), while the third is not depicted at all since it is too far away from the others.

\subsection{Analysis of the TTV Diagram}\label{ttvAnalysis}	

\begin{figure*}[ht]
\begin{center}$
\begin{array}{cc}
\includegraphics[width=16cm,height=9cm]{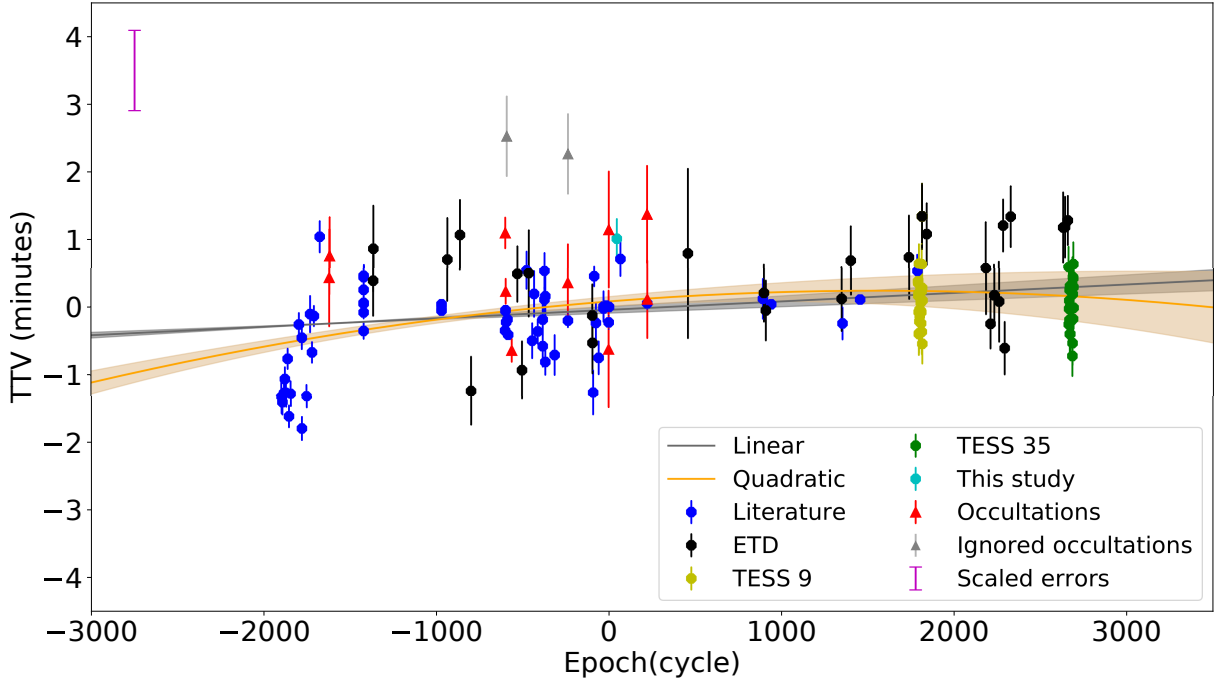}
\end{array}$
\end{center}
\caption{The TTV diagram of WASP-43\,b with linear (gray line) and quadratic (orange curve) models superimposed on the data. The red triangle marker shows occultations, the rest of the points were derived from transits. TESS data and our observation were marked according to the legend. Shaded areas represent the intervals between $16^{\rm th}$ and $84^{\rm th}$ percentile values for model parameters whereas the curves are for their median values. The magenta error bar on top left is to show the scale of the error bar assumed for all data points during both the linear and quadratic models.}
\label{fig:ttv}
\end{figure*}

We provided the resultant TTV diagram in Figure-\ref{fig:ttv} using the selected data set according to the reference elements including mid-transit (JD) $= 2457090.73888(8)$ stated in \cite{Weaver2020} and $P = 0.813474061(46)$ days \citep{Wong2020}. This is the most complete and homogenous TTV diagram to date, based on the most precise observations. We obtained the root means square (RMS) of the linear fit residuals as $\sim 38\; {\rm s}$, which is greater than the average standard errors of the observations ($18.3\; {\rm s}$).

To interpret the cause of these variations on the TTV diagram, we fitted the linear and quadratic models (Figure-\ref{fig:ttv}). Both the linear model and the linear term of the quadratic model originate from the accumulation of measurement errors in reference values for the mid-transit time and the orbital period with the number of transit epochs that have passed since reference ephemeris. On the other hand, the quadratic term can be caused by a potential secular variation in the orbital period if it is found to be statistically significant. 

We observed a scatter larger than the error bars for the entirety of the TTV diagram, including the occultations. However, the scatter is especially large compared to the observational uncertainties for the epochs $< -1752$ on the TTV diagram. This corresponds to the light curves published by \cite{Gillon2012}. We tried to figure out what was causing the problem. According to \cite{Gillon2012}, deduced parameters from their light curves reveal some extra scatter, which they attribute to the correlated noise of their data and, possibly, to the spot-crossing events during some transits. Activity-induced spots cause asymmetries in the light curves that lead to the incorrect description of the mid-transit times. The hypothesis becomes more substantial by the out-of-transit flux variability that we observed in TESS data (section-\ref{rotation}), which can be related to the host star rotation. However, if it was the only cause, then the scatter in the occultations larger than their error bars would not be observed, necessitating a careful look and explanation.

We conducted experiments to find out how timing errors affect the model parameter values and compared the findings in Table-\ref{table_tests}. These fits were done based on non-linear least-squares fitting using the SciPy package \citep{Virtanen2020}. First, we collected the data into three groups, the first group with error bars smaller than or equal to the average of error bars of the TESS data points ($19.87\; {\rm s}$) as test-1. The second group with error bars less than ($43.2\; {\rm s}$) (test-2), and the third group included all data points with error bars derived from their models (test-3). Test-4 is for all data points, while we scaled the error bars to $99.5\%$ confidence levels ($3\sigma$) for mid-transit times with the epochs $< -1752$, which exaggerates their influence on the models. The fifth test is for all data points, but this time, we scaled the error bars of the epochs $< -1752$ and the error bars of data points from light curves derived from transmission spectroscopy to 99.5\% confidence levels ($3\sigma$). Moreover, the sixth test is when we used $3\sigma$ for all points with small error bars ($ < 15.55\; {\rm s}$), to create a consistent data set. As the seventh test, we removed the data points with epochs $< -1752$ and scaled the error bars of the rest according to the variance around the linear model. Then we scaled all data point‘s error bars according to the variance around the linear model as the final test.
 
\begin{table*}[ht]
\caption{The experiments on error-bars in TTV diagram}    
\centering
\begin{center}
\footnotesize
\begin{tabular}{c c c c c c c c }
 \hline
 \hline
		Experiments & $\chi^2_{\nu,lin}$ & $\chi^2_{\nu,quad}$ & $BIC_{\nu,lin}$& $AIC_{\nu,lin}$& $BIC_{\nu,quad}$& $AIC_{\nu,quad}$& Quadratic coefficient \\
		\hline
		test 1& 13.69& 13.82&23.08 & 17.70&27.90 &19.82 & $(0.44 \pm 1.88) \times 10^{-11} $\\
		test 2& 10.83& 10.89& 20.79& 14.84& 25.82& 16.89& $(0.86 \pm 1.64) \times 10^{-11} $\\
		test 3& 10.52& 10.57& 20.54& 14.52& 25.60&16.57 & $(0.85 \pm 1.61) \times 10^{-11} $\\
		test 4& 8.11& 7.43& 18.13&12.11& 22.46& 13.42& $(5.87 \pm 1.53) \times 10^{-11} $\\
		test 5& 7.56& 6.46&17.58&11.56& 21.49& 12.46& $(8.75 \pm 1.70) \times 10^{-11} $\\
		test 6& 2.40& 2.41&12.42 & 6.40&17.45&8.42& $(-0.10 \pm 1.53) \times 10^{-11} $\\
		test 7& 0.99& 0.99& 10.83& 4.99&15.75&6.99& $(2.29 \pm 2.41) \times 10^{-11} $\\
		test 8& 0.99& 0.94& 11.01& 4.99&15.98&6.94& $(-5.09 \pm 1.79) \times 10^{-11} $\\

        \hline
\end{tabular}
\end{center}
\label{table_tests}
\end{table*}

 We made use of the Akaike Information Criterion (AIC) and the Bayesian Information Criterion (BIC) as well as examining the $\chi^2_{\nu}$ statistics for the model selection. The positive $\Delta BIC$ values are against the quadratic model, while $\Delta AIC$  values indicate that the quadratic model is almost as good as the linear model. 
 In all cases, the $\chi^2_{\nu,quad}$ of the quadratic fit is found close to the $\chi^2_{\nu,lin}$ of the linear fit. As we scaled the error bars within different groups $\chi^2_{\nu}$ is approached to 1.0, and when we assigned a single error bar to all data points equal to a 1-standard deviation of the scatter about the linear model, we get about 1.0 for the $\chi^{2}_{\nu}$. This convergence to 1 as we scaled the error bars of more and more data points hinting that the error bars are underestimated for some of the data points, or there is a real, short-term variability with a larger amplitude implied by the error bars. Hence, we searched for periodicity within a Lomb-Scargle periodogram  \citep{lomb1976, Scargle1982} in the TTV too. There are two peaks corresponding to $\sim 12.3$ and $\sim 50$ day-periodicities with FAP values larger than 30\%. We phased the TTV diagram accordingly and observed that the TTV data set does not agree with the sinusoidal-models based on them. This lack of statistically significant peaks at high-frequencies showed no evidence for a short-period variation in the TTV of the system given our data set. We used the Lomb-Scargle function in the timeseries module of the {\sc astropy} package \citep{VanderPlas2015} for the purpose. We also found a peak at a lower frequency, corresponding to $\sim 662$ days, with a lower FAP value (15.6\%). However, the considerable FAP value prevents a firm conclusion about its existence in the data for the moment.
 
 Consequently, we decided to continue with the test-8 approach and assign the same error bar, which is equal to the 1-standard deviation about a linear model, and weighted all data points equally for the rest of our analysis and fitted the sample with linear and quadratic models again within a probabilistic fitting scheme. We did not used the results of test-8, instead we sampled from the posterior probability distributions of the coefficients of both models (linear and quadratic) within 64 Markov Chain Monte Carlo (MCMC) walkers, each of which was run for 50000 iterations with the Pymc3 package in Python \citep{Salvatier2016} after discarding the first 1000 steps in each walker as the burn-in period of the MCMC. The BIC and AIC values favor the linear model $ \chi^2_{\nu} = 0.99$, $ BIC = 11.01$ and $AIC=4.99$ over the quadratic model $ \chi^2_{\nu} = 0.94$, $BIC = 15.98$ and $AIC=6.94$. While the positive $\Delta BIC$ value $(4.97)$ is against the quadratic model, $\Delta AIC = 1.95$ indicates that the quadratic model is almost as good as the linear model. Based on the linear fit parameters, we update the reference linear ephemeris as follows:

\begin{equation}
       T_{\rm c,(BJD_{TDB})} = (2457090.738886(34)) + ( 0.813474147(21))\times E (days)
\label{tc1}
\end{equation}

Although the linear model is statistically better, the quadratic model cannot be ignored. For this reason, we are going to discuss the results of the quadratic model in section-\ref{sect2}.

\subsection{Potential Orbital Decay in WASP-43\,b}\label{sect2}

WASP-43\,b is one of the best candidates for decay in its orbit to be observed due to its ultra-short orbital period, which is less than a day, and its high mass only very slightly less than $2.0$ M$_{\rm jup}$. The following equation can be used to derive the orbital period derivative ($\frac{dP}{dt}$) assuming circular orbit \citep{patra2017}

 \begin{equation}
    T_{mid}(E) = T_{0} + PE+\frac{1}{2}\frac{dP}{dt} E^{2}.
	\label{Tc2}
\end{equation}

The change in the orbital period derivative was found as $\frac{dP}{dt} = (-1.11\pm 0.21)\times10^{-10}$ from the quadratic coefficient of the best fitting parabola indicating a decrease in the orbital period with a rate of $-0.0035 \pm 0.0007$ seconds per year.  

Assuming that the planet’s mass stays constant \cite{Goldreich1996} stated that the decay rate is

 \begin{equation}
    \frac{dP}{dt}\approx -\frac{27\pi}{2Q_{* }^{'}} (\frac{R_{*}}{a})^{5}\frac{M_{P}}{M_{*}}.
	\label{dpdt}
\end{equation}

Using parameters derived from our global model $M_{P}(M_{J})=1.909 \pm 0.052$, $M_{*}(M_{\odot})=0.645_{-0.025}^{+0.026}$, $R_{*} (R_{\odot})=0.6489_{+0.0097}^{-0.0098}$ and $a(au)=0.01475_{-0.00020}^{+0.00019}$ from Equation-\ref{dpdt}, we calculated $Q_{* }^{'}=(4.01 \pm 1.15)\times 10^{5}$ for the lower limit of the host star’s tidal quality parameter. 

\subsection{A hint of a downward trend in the orbital period after two years of TESS observations}

Having not found strong evidence for orbital decay in our analysis of the TTV diagram, we attempted a comparison of orbital periods derived from from different ways. Including 1) Periodograms (Lomb-Scargle \citep{Scargle1982, lomb1976}, Plavchan \citep{Plavchan2008}, the Fourier analysis with Period04 \citep{Lenz2005}, and BLS \citep{Kovacs2002}) of each sector of observations with TESS, 2) best-fitting combined transit models in each of these sectors extracted using the TESS pipeline and {\sc exofast-v2}, which provide orbital periods determined from combined transit models within the least-squares minimization approach. Both attempts turned out to be inconclusive since the derived orbital periods were not precise enough for such a comparison in each case.   

\begin{figure}[ht]
\begin{center}
\includegraphics[width=10cm,height=4cm]{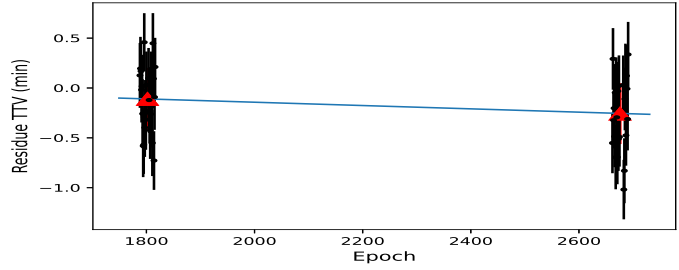}
\end{center}
\caption{A potential downward trend in the orbital period between two TESS observation, their averages represented by red triangles on TESS Clumps.}
\label{tesstrend}
\end{figure}

Using the corrected linear ephemeris (Equation-\ref{tc1}) we plotted the residuals of TTV for only the TESS observations versus the epoch of observation in Figure-\ref{tesstrend}. The corrected TTV for all mid-transit times of each sector was displayed with best-fitting lines to visualize a potential trend. The TESS transits in sector-35 have been observed earlier than in sector-09, best predicted by the new precise linear ephemeris. The slope of the line in Figure-\ref{tesstrend} ( $(-1.15 \pm 0.76) \times 10^{-7}$ day) indicates a decrease of the orbital period between sectors-09 \& 35 after a two-year interval and the timing difference between the averages of TESS clumps after ephemeris correction is $-0.15$ minute with the standard deviation of the averages as $0.3$ minute. Although the uncertainties of the results preclude a firm conclusion, the downward trend between two TESS clumps makes the target even more interesting for follow-up. 

\subsection{Spot induced modulations in the out-of-transit flux and rotation period of the host star}\label{rotation}

\begin{figure*}[ht]
\begin{center}$
\begin{array}{cc}
\includegraphics[width=8.0cm,height=6.5cm]{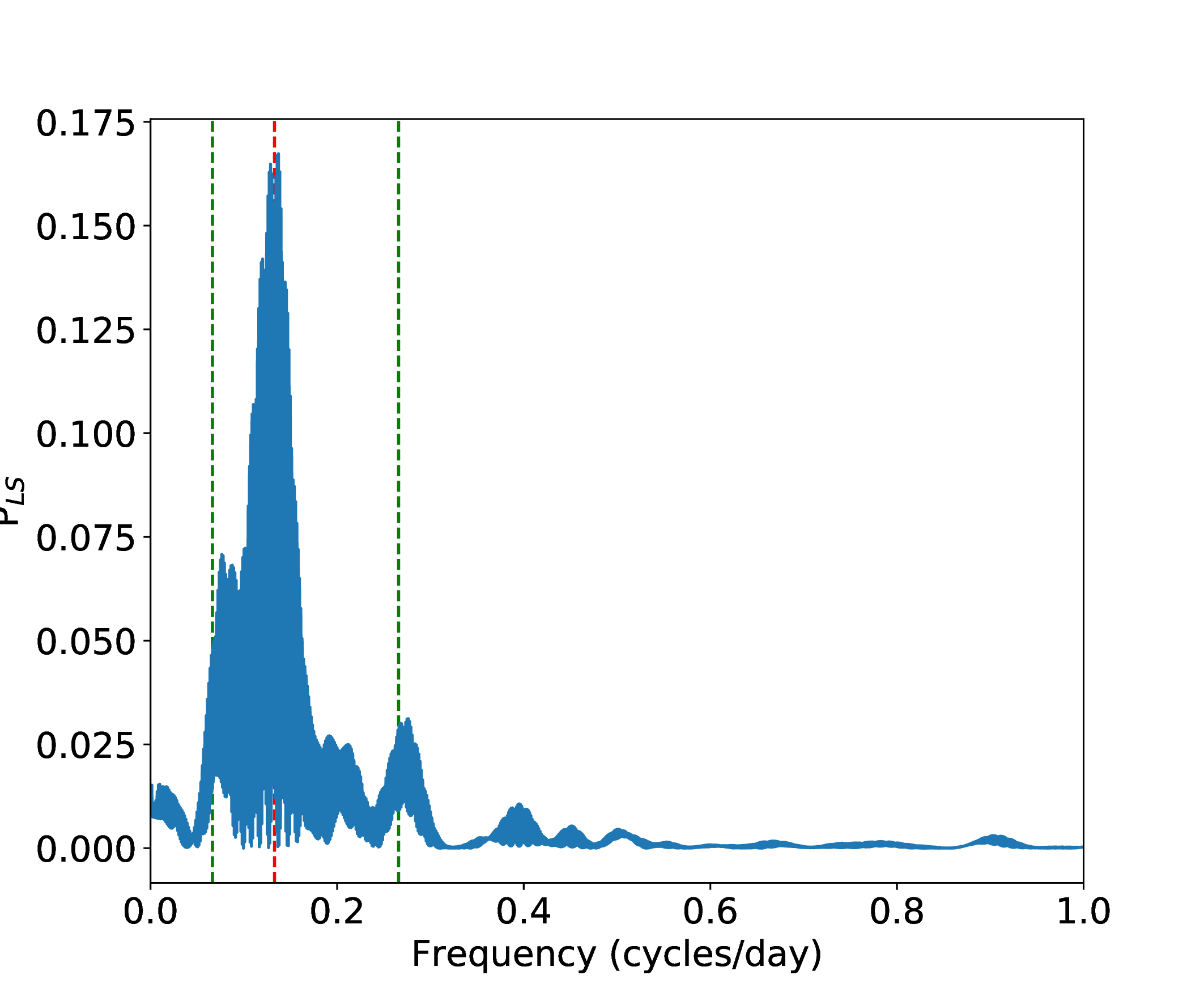}&
\includegraphics[width=10.4cm,height=6.5cm]{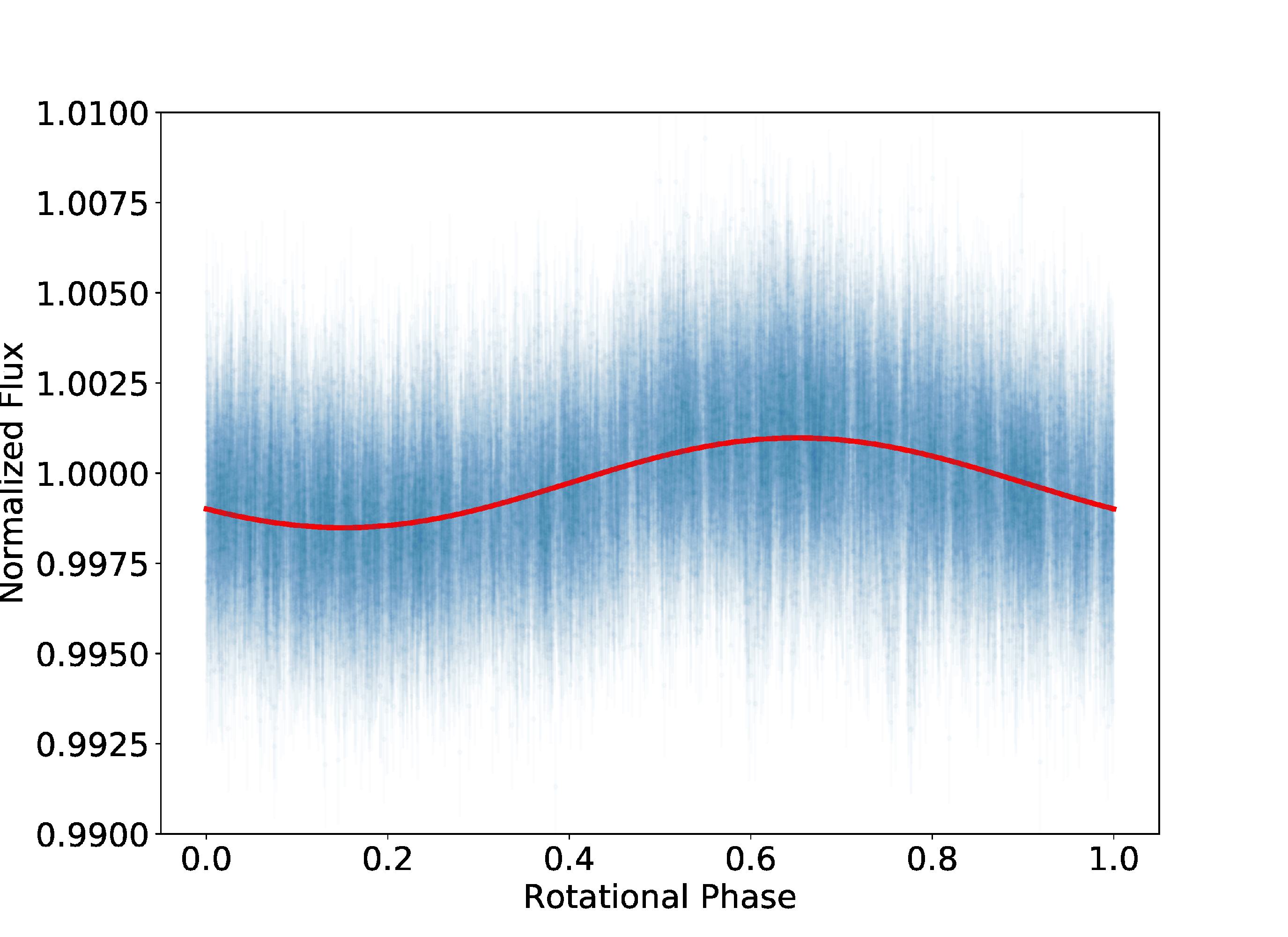}
\end{array}$
\end{center}
\caption{((Left panel) The periodogram of the combined light curve of the TESS sectors-09 \& 35. Red, dashed vertical line shows the position of the fundamental frequency ($f_{\rm max}$), green, dashed vertical lines on each side of it are for $f_{\rm max}$/2 and 2$f_{\rm max}$. (Right panel) The combined light curve (blue data points) and the sinusoidal model are based on the median values of the posterior distributions of related parameters phased with the derived rotation period of 7.52 days. The thickness of the curve is scaled to $1\sigma$-deviation.}
\label{outOfTransit}
\end{figure*}

WASP-43 displays strong sinusoidal variations in its out-of-transit fluxes in the light curves derived from observations in both TESS sectors-09 \& 35. Since the fluxes derived by the Simple Aperture Photometry of the SPOC pipeline (SAP fluxes) are subject to instrumental effects, we preferred to study the Pre-search Data Conditioning SAP fluxes (PDCSAP) \citep{Martins2020}. Out-of-transit variability is self-evident on the light curves from both sectors, which can be attributed to activity-induced surface inhomogeneities modulating the flux with the stellar rotation. Because the systematic effects should have already been removed from the SAP fluxes, reflection effect should be modulated with the orbital period, ellipsoidal and Doppler-boosting effects should have much lower amplitudes. We studied the dominant frequencies in these variations from both sectors based on the Lomb-Scargle periodograms, we calculated with the {\emph LombScargle} function \citep{VanderPlas2015} of the {\emph timeseries} module in the {\sc astropy} package \citep{Astropy2018}. We determined the frequencies of the maximum power in the Lomb-Scargle periodograms of this variation as 0.1312 day$^{-1}$ in sector-09 and 0.1348 day$^{-1}$ in sector-35, which correspond to periodicities of 7.62 and 7.42 days, respectively. We removed the data points at the beginning and the end of all gaps before a frequency analysis, since the CCD response changes dramatically at the time of data transmissions. When we combined the light curves from the two sectors and performed a frequency analysis on the combined light curve, we obtained the frequency at the maximum power as 0.1348 day$^{-1}$, which corresponds to a periodicity of 7.29 days. We then computed posterior probabilities of the parameters of a perfect sinusoidal, the prior values of which are set to the values from its Lomb-Scargle model, by sampling with a No-U-Turn Sampler (NUTS) run in 16 walkers of 5000 iterations each with the help of {\emph pymc3} code \citep{Salvatier2016}. The period derived from the out-of-transit variability is $ 7.5243 \pm 0.0004$ days, the value and the uncertainty of which depicts the median and $1\sigma$ of the sample from the posterior probability distribution, respectively. We provide the periodogram of the combined light curve and its best-fitting sine function superimposed on the light curve in Figure-\ref{outOfTransit}.

All the derived periodicities are very close to half of the value found as the stellar rotation period of WASP-43 by \cite{Hellier2011} as $15.6 \pm 0.4 $ days using a sine-wave fitting on the 2009 WASP-South data. However, TESS photometry is more precise compared to WASP photometry. Its sampling rate in the 2-minute cadence mode is also superior. In addition, TESS observed the target in two sectors continuously for more than 50 days. These, in turn, could have suppressed the fundamental frequency found by \cite{Hellier2011}, while strengthening the peak at $\sim7.3$ days, which they also found in their periodogram but discarded due to its lower amplitude. \cite{Martins2020} found a periodicity of 7.41 days for the star from the peak of the wavelet spectrum of the PDCSAP light curve derived from the sector-09 data. However, they listed it amongst other 32 stars with "dubious" rotation periods in their Table-2, but did not provide an uncertainty value. Assuming this variation is due to the spot-induced surface inhomogeneities modulated by the stellar rotation, we found the stellar rotation rate as $P_{rot} = 7.5243 \pm 0.0004 $ days from the combined PCDSAP light curves recorded in the sectors-09 \& 35. The uncertainty of the value only shows the success of the sinusoidal fit; hence it should be interpreted as the goodness of fit rather than the uncertainty of the suggested stellar rotation rate. As a result, we confirm the faster rotation rate of the host-star found by \cite{Martins2020} from sector-09 data and improve it with a more precise value thanks to the addition of the sector-35 data as $7.52$ days, with respect to which the entire out-of-transit light curve from sectors-09 \& 35 is phased in Figure-\ref{outOfTransit}. 

\section{Discussion}\label{sec:conc} 

We presented a global model of the transit light curves selected according to their quality and wavelength of acquisition, and precise RV data for WASP-43 system from which we derived the absolute parameters of the host star and the planet. Our fundamental RV and transit parameters are in agreement with previous studies except for the transit depth from \cite{Ricci2015}. Some of their observations seem to be influenced by weather conditions, affecting the transit depth in turn. We also found the host star WASP-43 to be significantly colder than the previous studies. Therefore we found smaller values for the stellar mass and the radius than suggested in most of the previous works on the system. Although our $T_{\rm eff}$ value is ~300 K colder than found by \cite{Hellier2011}, our stellar mass and radius values are larger compared to theirs. \cite{Hellier2011} pointed out that the method they employed does not constrain the stellar mass and radius in this mass regime satisfactorily. Therefore they derived alternative parameters with the infrared flux method, which agrees better with our results. \cite{Hellier2011} also emphasized the need for an update on the spectral analysis  with better data. \cite{Esposito2017} analyzed the weighted mean of high-quality HARPS-N spectra with the equivalent width (EW) method. However, the EW method is known to suffer from heavy line blending and low number statistics due to the lack of the sufficient number of doubly-ionized iron lines (FeII) in cold stars' spectra.

We performed a homogeneous analysis of the light curves of the WASP-43 system acquired by a decade of ground and space observations to create its TTV diagram. Mid-transit times across the two sectors (09 \& 35) of TESS observations, and a precise mid-transit time from our observation (with a meter-class telescope T100), are included in the TTV diagram too, spanning the largest time interval studied so far in the literature. We used the same analysis method to measure the mid-transit times from high-quality published and ETD light curves, as a result of which we homogenized the data set for a thorough TTV analysis. We selected the best light curves for this study and produced the most complete and precise TTV diagram.

We refined the reference linear ephemeris based on a linear fit to the data set. We used linear and quadratic models to interpret the cause of variations of the TTV diagram over ten years. According to the statistical results, the linear model turned out to be a better fit than the parabolic model, but the parabolic model cannot be overlooked. The TTV diagram reveals a tentative downward trend in the period between two sectors of TESS observations after the ephemeris correction. The negative quadratic coefficient of the best fitting parabola points to a decrease in the orbital period with the rate $\frac{dP}{dt} = -0.0035 \pm 0.0007 \; s \; yr^{-1}$. Therefore, a potential decay rate has to be smaller than $\dot{P}$ given our data, which implies that the star dissipates at least that amount of energy. Hence we found a lower limit for the tidal quality parameter as $Q_{\;*}^{'} > (4.01 \pm  1.15) \times 10^{5}$. This finding is in agreement with \cite{Hoyer2016}, because they claimed, after about five years, it would be possible to probe the $Q_{\;*}^{'}< 10^6$ regime. Moreover, \cite{Penev2018} suggested $Q_{\;*}^{'}$ for the stars hosting gas giant planets with the orbital periods between $0.5–2$ days to take values in the $10^5$ to $10^7$ range.
\\
The planet's remaining lifetime ($ \tau_{a}$) then can be calculated from the Equation-\ref{tau} \citep{Levrard2009}.

\begin{equation}
       \tau_{a} = \frac{P}{48} \frac{Q_{\;*}^{'}}{2\pi} \frac{M_{*}}{M_{P}} (\frac{a}{R_{*}})^{5}. 
\label{tau}
\end{equation}

We computed $\tau_{a} = 2.83 \pm 1.13$ Myr for WASP-43 b assuming derived $Q_{\;*}^{'}$ and the parameter values from our global model.  

Since the mid-occultation times in the TTV diagram follow the same pattern as the mid-transit times, apsidal motion is ruled out. For an orbit to be nearly circular ($e = 0.0035_{-0.0025}^{+0.0060}$ \cite{Gillon2012}), this is exactly what we would expect. The short baseline of observations still does not allow us to reach firm conclusions whether the orbit of this ultra-short period planet is decaying or not. 

We presented a model-independent rotational period for the host-star as 7.52 days assuming the observed variability in the out-of-transit fluxes in the TESS data is due to spot-induced light curve modulations with the stellar rotation. \cite{Hellier2011} measured the projected rotation rate $v~ sin i = 4.0 \pm 0.4$ km/s from high-resolution spectroscopy, which was not in agreement with the rotation period they derived from out-of-transit variability in WASP light curve as 15.6 days. Assuming the rotation rate they found is correct, the projected rotational velocity should have been half of the value they derived from spectroscopy. Our result from the superior TESS photometry for the rotation rate as 7.52 days is in perfect agreement with the projected rotational velocity value they found. Hence, it does not necessitate an additional broadening mechanism (e.g., a larger macroturbulent velocity) to explain the disagreement as they suggested. It also implies that the axes of stellar rotation and planetary orbit are aligned considering the uncertainties of the measurements.

WASP-43 b is an active star with log(R'$_{HK}$) = -4.35 \citep{Esposito2017}. Therefore one expects the star to be young. The age we found for the star from stellar isochrones is 8.5 Gyrs, but this technique is known not to work well for low mass stars because the fundamental parameters of these stars change very slowly in a very narrow range during their main-sequence evolution than the models can predict. We attempted to measure the star's age with the gyrochronology method by employing the {\sc TATOO} code \citep{florian2020} based on the formulation given by \cite{angus2015}. We found the star to be very young (~200 Myr) based on the stellar rotation (P$_{rot}$) value from our work. Note that \cite{Hellier2011} calculated the star's age as ~400 Myr with the same method based on their value of P$_{rot}$, which is twice the value we found from the out-of-transit variability in TESS photometry. Massive planets within the close vicinity of their stars such as WASP-43\,b can spin up their stars by transferring angular momentum through tidal interactions, causing an underestimation of the age through gyrochronology. This delay in the rotational evolution can be accounted for, interpolated based on the system parameters, and corrected with the {\sc TATOO} code to measure a "tidal-chronology age" of the star. However, unfortunately, we could not estimate it because WASP-43 is out of the range in which the model provides reliable results. As a result, it can be argued that the star is too young, potentially with 200 Myrs as a lower limit on its age. From the empirical log(R'$_{HK}$) - P$_{rot}$ correlation given by \cite{suarez2015} (Equation-9) log(R'$_{HK}$) value for WASP-43 can be computed as -4.35 $\pm 0.15$ based on the P$_{rot}$ value from our study which is in 1-$\sigma$ agreement with the measured log(R'$_{HK}$) = -4.25 $\pm 0.10$ \citep{Esposito2017}. With the rotation period from \cite{Hellier2011} as 15.6 days, the derived log(R'$_{HK}$) value as -4.62 also implies a less active star than the spectroscopic evidence suggests \citep{Esposito2017}. This disagreement is explained by the authors as the result of a probable magnetic interaction between the planet and the host star. The rotation rate we derived from TESS photometry does not require such an assumption as well, but it provides a clue that the star is younger than once thought.\\
The Transmission Spectroscopy Metric (TSM) for a planet was presented at \cite{Kempton2018} as

\begin{equation}
        \rm{TSM}= (\rm{sf}) \times \frac{R^{3}_{P}\; T_{eq}}{M_{P}\; R^{2}_{*}} \times 10^{\frac{-m_{J}}{5}},
\label{tsm}
\end{equation}

where sf is a scale factor, $T_{eq}$ and $m_{J}$ are the planet’s equilibrium temperature and the apparent magnitude of the host star in the J-band, respectively. We computed the $TSM = 83 \pm 12$ for WASP-43\,b assuming its radius is near the upper limit for sub-Jovians $(10.424_{-1.009}^{+0.785}\, M\earth$, \cite{Hellier2011}); considering $1.15$ as the scale factor (as given in Table-1 of \cite{Kempton2018} to normalization) and $9.995$ as its $m_{J}$ \citep{2003tmc..book.....C}. This makes the target even more interesting in terms of studying its atmosphere through transmission spectroscopy because the derived TSM value ($83 \pm 12$, this study) is in agreement with the lower limit for the TSM value (90) required for JWST observations for sub-Jovians \citep{Kempton2018}, for which it is already in the prime list of targets \citep{Moliere2017}.
\\
The mass-loss rate resulting from the total X-ray and Extreme Ultra-violet (EUV) radiative flux ($F_{XUV} = F_{X} + F_{EUV}$) at the planet orbit impinging on the atmosphere of the planet WASP-43\,b can be calculated using the equation-\ref{mdot} \citep{Salz2015, Sanz2011}

\begin{equation}
       \dot{M} = \frac{3\eta F_{XUV}}{4KG \rho_{p}}.
\label{mdot}
\end{equation}

\begin{figure}[ht]
\begin{center}
\includegraphics[width=12cm,height=8cm]{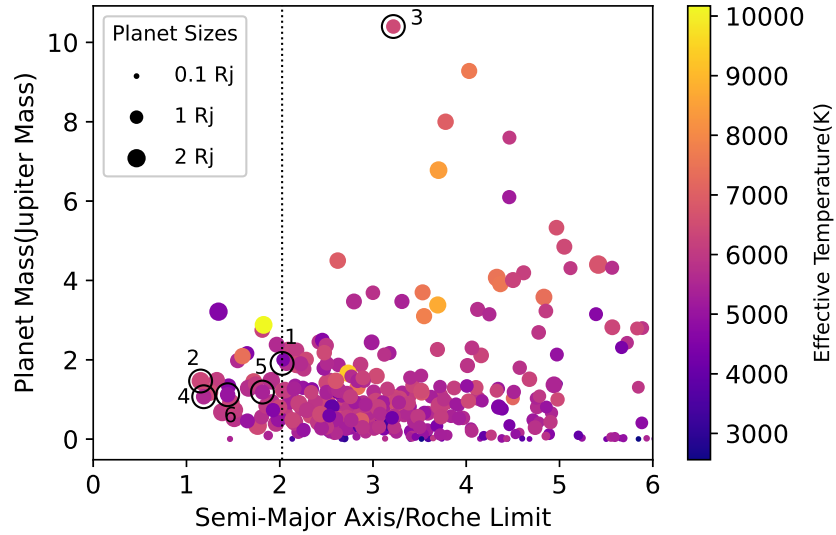}
\caption{Mass versus $\frac{a}{a_{R}}$ for over 400 hot-Jupiters. The size of the markers for each planet is scaled to its radius. The color scale displays the temperature of their host stars. Some ultra-short period hot-Jupiters such as WASP-43\,b, WASP-12\,b, WASP-18\,b, WASP-19\,b, WASP-4\,b, and WTS-2\,b are labeled with the numbers of 1, 2, 3, 4, 5, and 6 respectively.}
\label{fig:aar}
\end{center}
\end{figure}

Where $G$ is the gravitational constant, $\rho_{p}$ is the density of the planet, $\eta$ is the heating efficiency, and K is the potential energy reduction factor due to the stellar tidal forces that $K\leq 1$ \citep{Erkaev2007}. Assuming $K=1$, $\eta = 0.15$ and $\rm F_{\rm XUV} = 6.6 \times 10^{4}$ erg cm$^{-2}$ s$^{-1}$ as given in \cite{Salz2015} and using the parameter values derived from our global model, we measured $\log \dot{\rm M} = 11.26 \pm 9.81$ g/s as a lower limit to the mass loss of the atmosphere of WASP-43\,b. This is in agreement with $ \log \dot{\rm M} = 10.48$ g/s presented in \cite{Salz2015}.
\\
This estimated mass loss due to a hydrodynamic planetary wind \citep{Salz2015} will need to be confirmed by observable evidence by examining its atmosphere for evidence such as absorption features in the metastable Helium (He*) lines which were found for WASP-69\,b \citep{Lile2020}. Since TSM ($83 \pm 12$) is proportional to the Signal-to-Noise Ratio (SNR) of the predicted transmission spectrum \citep{Kempton2018}, the planet can be an excellent candidate for atmospheric observations with The James Webb Space Telescope (JWST) \citep{Kempton2018} to investigate a potential mass loss from its atmosphere.
\\
There is a decrease in the number of short-period hot-Jupiters around 2 Roche-radii ($a / a_{\rm R}=2$), the radius at which planet's eccentric orbit tends to circularize \citep{Ford2006}, when $\sim 400$ hot-Jupiters listed in the NASA Exoplanet Archive with precise mass and radii are considered.  Only 4\% of the 231 transiting giant planets studied by \cite{Bonomo2017} were found to be in $a / a_{\rm R} < 2$. Such hot-Jupiters are even more rarely found around low-mass stars. Because of their small orbital separations, their lifetimes are shortened due to tidal decay \citep{Hellier2011}. Figure-\ref{fig:aar} depicts a plot of hot-Jupiter mass versus the semi-major axis to Roche radii ratio $a / a_{\rm R}$, with the radii and color-scale that reflects the temperature of their host stars. In order to construct this plot we used the NASA Exoplanet Archive \footnote{https://exoplanetarchive.ipac.caltech.edu/} on March 1, 2021, with the following criteria: $0 <  a < 0.06$ au, $0.01 < M_{\rm P} < 12 \; M_{\rm jup}$, $0 < R_{\rm P} < 2 \; R_{\rm jup}$, and  $0 < M_{\star} < 2 \; M _{\odot}$. The Roche radii were then calculated using the Equation-\ref{roche_radius} \citep{Ford2006}.

\begin{equation}{
a_{R }=\frac {R_{P}}{0.462 (\frac{M_{P}}{M_{*}} )^{1/3}}
}
\label{roche_radius}
\end{equation}

Figure-\ref{fig:aar} has more short-period hot-Jupiters in a / a$_{\rm R} < 2$ than the study by \cite{Bonomo2017} since more planets should have been discovered and the orbital parameters of the already-known planets might have been refined in the meantime. We measured the correlation coefficients between the plot's parameters. There is a moderate correlation $(r = 0.64)$ between $T_{\rm eff}$ and $R_{\rm P}$. According to this correlation, higher-mass hot-Jupiters orbit hotter stars, while lower mass hot-Jupiters tend to orbit cooler stars. WASP-43\,b is one of the most massive planets amongst others orbiting a cold star, and it is located at the boundary of $a / a_{\rm R}=2$, which we depicted with a dashed vertical line on the plot.

\vspace{1.5cm}
\acknowledgments
{We thank T\"UB{\.I}TAK National Observatory of Turkey (TUG) for partial support in using the T100 telescope with project number 12CT100-378. O.B. gratefully acknowledges the support by The Scientific and Technological Council of Turkey (T\"UB{\.I}TAK) through the research grand 118F042.This work has used data from NASA’s Transiting Exoplanet Survey Satellite-TESS exoplanet mission (\it {https://tess.mit.edu/}). We thank the observers (J. Gonzalez; E. Nicolas; J. L. de Haro; P. Evans; P. Benni; G. Marino; C. Andre; J. Clement; J. Nougayrede; E. D. Alonso; A. Valdera; A. Ravenna; Y. Jongen; N. Montigiani; M. Mannucci; F. Scaggiante; D. Zardin; M. Fiaschi; M. Bretton) and the organizers of the Astronomical Society of the Czech Republic, the Exoplanet Transit Database (ETD) section for transit light curves collected (\it {http://var2.astro.cz}).
}

\software{AstroImageJ \citep{Collins2017}, lightkurve \citep{Cardoso2018}, ASTROPY \citep{Astropy2018, VanderPlas2015}, SciPy \citep{Virtanen2020}, Pymc3 \citep{Salvatier2016}, EXOFAST-V1 \citep{Eastman2013}, EXOFAST-V2 \citep{eastman2019}, TATOO \citep{florian2020}, LMFIT \citep{Newville2016}.}
\clearpage
 
\bibliographystyle{aasjournal}
\bibliography{new.ms}

\begin{thebibliography}{}
\expandafter\ifx\csname natexlab\endcsname\relax\def\natexlab#1{#1}\fi
\providecommand{\url}[1]{\href{#1}{#1}}
\providecommand{\dodoi}[1]{doi:~\href{http://doi.org/#1}{\nolinkurl{#1}}}
\providecommand{\doeprint}[1]{\href{http://ascl.net/#1}{\nolinkurl{http://ascl.net/#1}}}
\providecommand{\doarXiv}[1]{\href{https://arxiv.org/abs/#1}{\nolinkurl{https://arxiv.org/abs/#1}}}

\bibitem[{{Angus} {et~al.}(2015){Angus}, {Aigrain}, {Foreman-Mackey}, \&
  {McQuillan}}]{angus2015}
{Angus}, R., {Aigrain}, S., {Foreman-Mackey}, D., \& {McQuillan}, A. 2015,
  \mnras, 450, 1787, \dodoi{10.1093/mnras/stv423}

\bibitem[{{Astropy Collaboration} {et~al.}(2018){Astropy Collaboration},
  {Price-Whelan}, {Sip{\H{o}}cz}, {G{\"u}nther}, {Lim}, {Crawford}, {Conseil},
  {Shupe}, {Craig}, {Dencheva}, {Ginsburg}, {VanderPlas}, {Bradley},
  {P{\'e}rez-Su{\'a}rez}, {de Val-Borro}, {Aldcroft}, {Cruz}, {Robitaille},
  {Tollerud}, {Ardelean}, {Babej}, {Bach}, {Bachetti}, {Bakanov}, {Bamford},
  {Barentsen}, {Barmby}, {Baumbach}, {Berry}, {Biscani}, {Boquien}, {Bostroem},
  {Bouma}, {Brammer}, {Bray}, {Breytenbach}, {Buddelmeijer}, {Burke},
  {Calderone}, {Cano Rodr{\'\i}guez}, {Cara}, {Cardoso}, {Cheedella}, {Copin},
  {Corrales}, {Crichton}, {D'Avella}, {Deil}, {Depagne}, {Dietrich}, {Donath},
  {Droettboom}, {Earl}, {Erben}, {Fabbro}, {Ferreira}, {Finethy}, {Fox},
  {Garrison}, {Gibbons}, {Goldstein}, {Gommers}, {Greco}, {Greenfield},
  {Groener}, {Grollier}, {Hagen}, {Hirst}, {Homeier}, {Horton}, {Hosseinzadeh},
  {Hu}, {Hunkeler}, {Ivezi{\'c}}, {Jain}, {Jenness}, {Kanarek}, {Kendrew},
  {Kern}, {Kerzendorf}, {Khvalko}, {King}, {Kirkby}, {Kulkarni}, {Kumar},
  {Lee}, {Lenz}, {Littlefair}, {Ma}, {Macleod}, {Mastropietro}, {McCully},
  {Montagnac}, {Morris}, {Mueller}, {Mumford}, {Muna}, {Murphy}, {Nelson},
  {Nguyen}, {Ninan}, {N{\"o}the}, {Ogaz}, {Oh}, {Parejko}, {Parley}, {Pascual},
  {Patil}, {Patil}, {Plunkett}, {Prochaska}, {Rastogi}, {Reddy Janga},
  {Sabater}, {Sakurikar}, {Seifert}, {Sherbert}, {Sherwood-Taylor}, {Shih},
  {Sick}, {Silbiger}, {Singanamalla}, {Singer}, {Sladen}, {Sooley},
  {Sornarajah}, {Streicher}, {Teuben}, {Thomas}, {Tremblay}, {Turner},
  {Terr{\'o}n}, {van Kerkwijk}, {de la Vega}, {Watkins}, {Weaver}, {Whitmore},
  {Woillez}, {Zabalza}, \& {Astropy Contributors}}]{Astropy2018}
{Astropy Collaboration}, {Price-Whelan}, A.~M., {Sip{\H{o}}cz}, B.~M., {et~al.}
  2018, \aj, 156, 123, \dodoi{10.3847/1538-3881/aabc4f}

\bibitem[{{Ba{\c{s}}t{\"u}rk} {et~al.}(2015){Ba{\c{s}}t{\"u}rk}, {Hinse},
  {{\"O}zavc{\i}}, {Y{\"o}r{\"u}ko{\v{g}}lu}, \& {Selam}}]{Basturk2015}
{Ba{\c{s}}t{\"u}rk}, {\"O}., {Hinse}, T.~C., {{\"O}zavc{\i}}, {\.I}.,
  {Y{\"o}r{\"u}ko{\v{g}}lu}, O., \& {Selam}, S.~O. 2015, in Astronomical
  Society of the Pacific Conference Series, Vol. 496, Living Together: Planets,
  Host Stars and Binaries, ed. S.~M. {Rucinski}, G.~{Torres}, \& M.~{Zejda},
  370.
\newblock \doarXiv{1508.04500}

\bibitem[{{Ba{\c{s}}t{\"u}rk} {et~al.}(2020){Ba{\c{s}}t{\"u}rk},
  {Yal{\c{c}}{\i}nkaya}, {Esmer}, {Tanr{\i}verdi}, {Mancini}, {Daylan},
  {Southworth}, \& {Keten}}]{basturk2020}
{Ba{\c{s}}t{\"u}rk}, {\"O}., {Yal{\c{c}}{\i}nkaya}, S., {Esmer}, E.~M.,
  {et~al.} 2020, \mnras, 496, 4174, \dodoi{10.1093/mnras/staa1758}

\bibitem[{{Barker} \& {Ogilvie}(2009)}]{Barker2009}
{Barker}, A.~J., \& {Ogilvie}, G.~I. 2009, \mnras, 395, 2268,
  \dodoi{10.1111/j.1365-2966.2009.14694.x}

\bibitem[{{Bianchi} {et~al.}(2017){Bianchi}, {Shiao}, \&
  {Thilker}}]{bianchi2017}
{Bianchi}, L., {Shiao}, B., \& {Thilker}, D. 2017, \apjs, 230, 24,
  \dodoi{10.3847/1538-4365/aa7053}

\bibitem[{{Birkby} {et~al.}(2014){Birkby}, {Cappetta}, {Cruz}, {Koppenhoefer},
  {Ivanyuk}, {Mustill}, {Hodgkin}, {Pinfield}, {Sip{\H{o}}cz}, {Kov{\'a}cs},
  {Saglia}, {Pavlenko}, {Barrado}, {Bayo}, {Campbell}, {Catalan}, {Fossati},
  {G{\'a}lvez-Ortiz}, {Kenworthy}, {Lillo-Box}, {Mart{\'\i}n}, {Mislis}, {de
  Mooij}, {Nefs}, {Snellen}, {Stoev}, {Zendejas}, {del Burgo}, {Barnes},
  {Goulding}, {Haswell}, {Kuznetsov}, {Lodieu}, {Murgas}, {Palle}, {Solano},
  {Steele}, \& {Tata}}]{Birkby2014}
{Birkby}, J.~L., {Cappetta}, M., {Cruz}, P., {et~al.} 2014, \mnras, 440, 1470,
  \dodoi{10.1093/mnras/stu343}

\bibitem[{{Blecic} {et~al.}(2014){Blecic}, {Harrington}, {Madhusudhan},
  {Stevenson}, {Hardy}, {Cubillos}, {Hardin}, {Bowman}, {Nymeyer}, {Anderson},
  {Hellier}, {Smith}, \& {Collier Cameron}}]{Blecic2014}
{Blecic}, J., {Harrington}, J., {Madhusudhan}, N., {et~al.} 2014, \apj, 781,
  116, \dodoi{10.1088/0004-637X/781/2/116}

\bibitem[{{Bonomo} {et~al.}(2017){Bonomo}, {Desidera}, {Benatti}, {Borsa},
  {Crespi}, {Damasso}, {Lanza}, {Sozzetti}, {Lodato}, {Marzari}, {Boccato},
  {Claudi}, {Cosentino}, {Covino}, {Gratton}, {Maggio}, {Micela}, {Molinari},
  {Pagano}, {Piotto}, {Poretti}, {Smareglia}, {Affer}, {Biazzo}, {Bignamini},
  {Esposito}, {Giacobbe}, {H{\'e}brard}, {Malavolta}, {Maldonado}, {Mancini},
  {Martinez Fiorenzano}, {Masiero}, {Nascimbeni}, {Pedani}, {Rainer}, \&
  {Scandariato}}]{Bonomo2017}
{Bonomo}, A.~S., {Desidera}, S., {Benatti}, S., {et~al.} 2017, \aap, 602, A107,
  \dodoi{10.1051/0004-6361/201629882}

\bibitem[{{Chambers} {et~al.}(2016){Chambers}, {Magnier}, {Metcalfe},
  {Flewelling}, {Huber}, {Waters}, {Denneau}, {Draper}, {Farrow}, {Finkbeiner},
  {Holmberg}, {Koppenhoefer}, {Price}, {Rest}, {Saglia}, {Schlafly}, {Smartt},
  {Sweeney}, {Wainscoat}, {Burgett}, {Chastel}, {Grav}, {Heasley}, {Hodapp},
  {Jedicke}, {Kaiser}, {Kudritzki}, {Luppino}, {Lupton}, {Monet}, {Morgan},
  {Onaka}, {Shiao}, {Stubbs}, {Tonry}, {White}, {Ba{\~n}ados}, {Bell},
  {Bender}, {Bernard}, {Boegner}, {Boffi}, {Botticella}, {Calamida},
  {Casertano}, {Chen}, {Chen}, {Cole}, {Deacon}, {Frenk}, {Fitzsimmons},
  {Gezari}, {Gibbs}, {Goessl}, {Goggia}, {Gourgue}, {Goldman}, {Grant},
  {Grebel}, {Hambly}, {Hasinger}, {Heavens}, {Heckman}, {Henderson}, {Henning},
  {Holman}, {Hopp}, {Ip}, {Isani}, {Jackson}, {Keyes}, {Koekemoer}, {Kotak},
  {Le}, {Liska}, {Long}, {Lucey}, {Liu}, {Martin}, {Masci}, {McLean}, {Mindel},
  {Misra}, {Morganson}, {Murphy}, {Obaika}, {Narayan}, {Nieto-Santisteban},
  {Norberg}, {Peacock}, {Pier}, {Postman}, {Primak}, {Rae}, {Rai}, {Riess},
  {Riffeser}, {Rix}, {R{\"o}ser}, {Russel}, {Rutz}, {Schilbach}, {Schultz},
  {Scolnic}, {Strolger}, {Szalay}, {Seitz}, {Small}, {Smith}, {Soderblom},
  {Taylor}, {Thomson}, {Taylor}, {Thakar}, {Thiel}, {Thilker}, {Unger},
  {Urata}, {Valenti}, {Wagner}, {Walder}, {Walter}, {Watters}, {Werner},
  {Wood-Vasey}, \& {Wyse}}]{chambers2016}
{Chambers}, K.~C., {Magnier}, E.~A., {Metcalfe}, N., {et~al.} 2016, arXiv
  e-prints, arXiv:1612.05560.
\newblock \doarXiv{1612.05560}

\bibitem[{{Chen} {et~al.}(2014){Chen}, {van Boekel}, {Wang}, {Nikolov},
  {Fortney}, {Seemann}, {Wang}, {Mancini}, \& {Henning}}]{Chen2014}
{Chen}, G., {van Boekel}, R., {Wang}, H., {et~al.} 2014, \aap, 563, A40,
  \dodoi{10.1051/0004-6361/201322740}

\bibitem[{{Choi} {et~al.}(2016){Choi}, {Dotter}, {Conroy}, {Cantiello},
  {Paxton}, \& {Johnson}}]{choi2016}
{Choi}, J., {Dotter}, A., {Conroy}, C., {et~al.} 2016, \apj, 823, 102,
  \dodoi{10.3847/0004-637X/823/2/102}

\bibitem[{{Collins} {et~al.}(2017){Collins}, {Kielkopf}, {Stassun}, \&
  {Hessman}}]{Collins2017}
{Collins}, K.~A., {Kielkopf}, J.~F., {Stassun}, K.~G., \& {Hessman}, F.~V.
  2017, \aj, 153, 77, \dodoi{10.3847/1538-3881/153/2/77}

\bibitem[{{Cutri} {et~al.}(2003){Cutri}, {Skrutskie}, {van Dyk}, {Beichman},
  {Carpenter}, {Chester}, {Cambresy}, {Evans}, {Fowler}, {Gizis}, {Howard},
  {Huchra}, {Jarrett}, {Kopan}, {Kirkpatrick}, {Light}, {Marsh}, {McCallon},
  {Schneider}, {Stiening}, {Sykes}, {Weinberg}, {Wheaton}, {Wheelock}, \&
  {Zacarias}}]{2003tmc..book.....C}
{Cutri}, R.~M., {Skrutskie}, M.~F., {van Dyk}, S., {et~al.} 2003, {2MASS All
  Sky Catalog of point sources.}

\bibitem[{{Cutri} {et~al.}(2021){Cutri}, {Wright}, {Conrow}, {Fowler},
  {Eisenhardt}, {Grillmair}, {Kirkpatrick}, {Masci}, {McCallon}, {Wheelock},
  {Fajardo-Acosta}, {Yan}, {Benford}, {Harbut}, {Jarrett}, {Lake}, {Leisawitz},
  {Ressler}, {Stanford}, {Tsai}, {Liu}, {Helou}, {Mainzer}, {Gettngs},
  {Gonzalez}, {Hoffman}, {Marsh}, {Padgett}, {Skrutskie}, {Beck}, {Papin}, \&
  {Wittman}}]{cutri2013}
{Cutri}, R.~M., {Wright}, E.~L., {Conrow}, T., {et~al.} 2021, VizieR Online
  Data Catalog, II/328

\bibitem[{{Eastman} {et~al.}(2013){Eastman}, {Gaudi}, \& {Agol}}]{Eastman2013}
{Eastman}, J., {Gaudi}, B.~S., \& {Agol}, E. 2013, \pasp, 125, 83,
  \dodoi{10.1086/669497}

\bibitem[{{Eastman} {et~al.}(2019){Eastman}, {Rodriguez}, {Agol}, {Stassun},
  {Beatty}, {Vanderburg}, {Gaudi}, {Collins}, \& {Luger}}]{eastman2019}
{Eastman}, J.~D., {Rodriguez}, J.~E., {Agol}, E., {et~al.} 2019, arXiv
  e-prints, arXiv:1907.09480.
\newblock \doarXiv{1907.09480}

\bibitem[{{Erkaev} {et~al.}(2007){Erkaev}, {Kulikov}, {Lammer}, {Selsis},
  {Langmayr}, {Jaritz}, \& {Biernat}}]{Erkaev2007}
{Erkaev}, N.~V., {Kulikov}, Y.~N., {Lammer}, H., {et~al.} 2007, \aap, 472, 329,
  \dodoi{10.1051/0004-6361:20066929}

\bibitem[{{Esposito} {et~al.}(2017){Esposito}, {Covino}, {Desidera}, {Mancini},
  {Nascimbeni}, {Zanmar Sanchez}, {Biazzo}, {Lanza}, {Leto}, {Southworth},
  {Bonomo}, {Su{\'a}rez Mascare{\~n}o}, {Boccato}, {Cosentino}, {Claudi},
  {Gratton}, {Maggio}, {Micela}, {Molinari}, {Pagano}, {Piotto}, {Poretti},
  {Smareglia}, {Sozzetti}, {Affer}, {Anderson}, {Andreuzzi}, {Benatti},
  {Bignamini}, {Borsa}, {Borsato}, {Ciceri}, {Damasso}, {di Fabrizio},
  {Giacobbe}, {Granata}, {Harutyunyan}, {Henning}, {Malavolta}, {Maldonado},
  {Martinez Fiorenzano}, {Masiero}, {Molaro}, {Molinaro}, {Pedani}, {Rainer},
  {Scandariato}, \& {Turner}}]{Esposito2017}
{Esposito}, M., {Covino}, E., {Desidera}, S., {et~al.} 2017, \aap, 601, A53,
  \dodoi{10.1051/0004-6361/201629720}

\bibitem[{{Ford} \& {Rasio}(2006)}]{Ford2006}
{Ford}, E.~B., \& {Rasio}, F.~A. 2006, \apjl, 638, L45, \dodoi{10.1086/500734}

\bibitem[{{Fulton} {et~al.}(2011){Fulton}, {Shporer}, {Winn}, {Holman},
  {P{\'a}l}, \& {Gazak}}]{fulton2011}
{Fulton}, B.~J., {Shporer}, A., {Winn}, J.~N., {et~al.} 2011, \aj, 142, 84,
  \dodoi{10.1088/0004-6256/142/3/84}

\bibitem[{{Gaia Collaboration} {et~al.}(2016){Gaia Collaboration}, {Prusti},
  {de Bruijne}, {Brown}, {Vallenari}, {Babusiaux}, {Bailer-Jones}, {Bastian},
  {Biermann}, {Evans}, {Eyer}, {Jansen}, {Jordi}, {Klioner}, {Lammers},
  {Lindegren}, {Luri}, {Mignard}, {Milligan}, {Panem}, {Poinsignon},
  {Pourbaix}, {Randich}, {Sarri}, {Sartoretti}, {Siddiqui}, {Soubiran},
  {Valette}, {van Leeuwen}, {Walton}, {Aerts}, {Arenou}, {Cropper}, {Drimmel},
  {H{\o}g}, {Katz}, {Lattanzi}, {O'Mullane}, {Grebel}, {Holland}, {Huc},
  {Passot}, {Bramante}, {Cacciari}, {Casta{\~n}eda}, {Chaoul}, {Cheek}, {De
  Angeli}, {Fabricius}, {Guerra}, {Hern{\'a}ndez}, {Jean-Antoine-Piccolo},
  {Masana}, {Messineo}, {Mowlavi}, {Nienartowicz}, {Ord{\'o}{\~n}ez-Blanco},
  {Panuzzo}, {Portell}, {Richards}, {Riello}, {Seabroke}, {Tanga},
  {Th{\'e}venin}, {Torra}, {Els}, {Gracia-Abril}, {Comoretto},
  {Garcia-Reinaldos}, {Lock}, {Mercier}, {Altmann}, {Andrae}, {Astraatmadja},
  {Bellas-Velidis}, {Benson}, {Berthier}, {Blomme}, {Busso}, {Carry},
  {Cellino}, {Clementini}, {Cowell}, {Creevey}, {Cuypers}, {Davidson}, {De
  Ridder}, {de Torres}, {Delchambre}, {Dell'Oro}, {Ducourant}, {Fr{\'e}mat},
  {Garc{\'\i}a-Torres}, {Gosset}, {Halbwachs}, {Hambly}, {Harrison}, {Hauser},
  {Hestroffer}, {Hodgkin}, {Huckle}, {Hutton}, {Jasniewicz}, {Jordan},
  {Kontizas}, {Korn}, {Lanzafame}, {Manteiga}, {Moitinho}, {Muinonen},
  {Osinde}, {Pancino}, {Pauwels}, {Petit}, {Recio-Blanco}, {Robin}, {Sarro},
  {Siopis}, {Smith}, {Smith}, {Sozzetti}, {Thuillot}, {van Reeven}, {Viala},
  {Abbas}, {Abreu Aramburu}, {Accart}, {Aguado}, {Allan}, {Allasia},
  {Altavilla}, {{\'A}lvarez}, {Alves}, {Anderson}, {Andrei}, {Anglada Varela},
  {Antiche}, {Antoja}, {Ant{\'o}n}, {Arcay}, {Atzei}, {Ayache}, {Bach},
  {Baker}, {Balaguer-N{\'u}{\~n}ez}, {Barache}, {Barata}, {Barbier}, {Barblan},
  {Baroni}, {Barrado y Navascu{\'e}s}, {Barros}, {Barstow}, {Becciani},
  {Bellazzini}, {Bellei}, {Bello Garc{\'\i}a}, {Belokurov}, {Bendjoya},
  {Berihuete}, {Bianchi}, {Bienaym{\'e}}, {Billebaud}, {Blagorodnova},
  {Blanco-Cuaresma}, {Boch}, {Bombrun}, {Borrachero}, {Bouquillon}, {Bourda},
  {Bouy}, {Bragaglia}, {Breddels}, {Brouillet}, {Br{\"u}semeister},
  {Bucciarelli}, {Budnik}, {Burgess}, {Burgon}, {Burlacu}, {Busonero}, {Buzzi},
  {Caffau}, {Cambras}, {Campbell}, {Cancelliere}, {Cantat-Gaudin}, {Carlucci},
  {Carrasco}, {Castellani}, {Charlot}, {Charnas}, {Charvet}, {Chassat},
  {Chiavassa}, {Clotet}, {Cocozza}, {Collins}, {Collins}, {Costigan}, {Crifo},
  {Cross}, {Crosta}, {Crowley}, {Dafonte}, {Damerdji}, {Dapergolas}, {David},
  {David}, {De Cat}, {de Felice}, {de Laverny}, {De Luise}, {De March}, {de
  Martino}, {de Souza}, {Debosscher}, {del Pozo}, {Delbo}, {Delgado},
  {Delgado}, {di Marco}, {Di Matteo}, {Diakite}, {Distefano}, {Dolding}, {Dos
  Anjos}, {Drazinos}, {Dur{\'a}n}, {Dzigan}, {Ecale}, {Edvardsson}, {Enke},
  {Erdmann}, {Escolar}, {Espina}, {Evans}, {Eynard Bontemps}, {Fabre},
  {Fabrizio}, {Faigler}, {Falc{\~a}o}, {Farr{\`a}s Casas}, {Faye}, {Federici},
  {Fedorets}, {Fern{\'a}ndez-Hern{\'a}ndez}, {Fernique}, {Fienga}, {Figueras},
  {Filippi}, {Findeisen}, {Fonti}, {Fouesneau}, {Fraile}, {Fraser}, {Fuchs},
  {Furnell}, {Gai}, {Galleti}, {Galluccio}, {Garabato}, {Garc{\'\i}a-Sedano},
  {Gar{\'e}}, {Garofalo}, {Garralda}, {Gavras}, {Gerssen}, {Geyer}, {Gilmore},
  {Girona}, {Giuffrida}, {Gomes}, {Gonz{\'a}lez-Marcos},
  {Gonz{\'a}lez-N{\'u}{\~n}ez}, {Gonz{\'a}lez-Vidal}, {Granvik}, {Guerrier},
  {Guillout}, {Guiraud}, {G{\'u}rpide}, {Guti{\'e}rrez-S{\'a}nchez}, {Guy},
  {Haigron}, {Hatzidimitriou}, {Haywood}, {Heiter}, {Helmi}, {Hobbs},
  {Hofmann}, {Holl}, {Holland}, {Hunt}, {Hypki}, {Icardi}, {Irwin}, {Jevardat
  de Fombelle}, {Jofr{\'e}}, {Jonker}, {Jorissen}, {Julbe}, {Karampelas},
  {Kochoska}, {Kohley}, {Kolenberg}, {Kontizas}, {Koposov}, {Kordopatis},
  {Koubsky}, {Kowalczyk}, {Krone-Martins}, {Kudryashova}, {Kull}, {Bachchan},
  {Lacoste-Seris}, {Lanza}, {Lavigne}, {Le Poncin-Lafitte}, {Lebreton},
  {Lebzelter}, {Leccia}, {Leclerc}, {Lecoeur-Taibi}, {Lemaitre}, {Lenhardt},
  {Leroux}, {Liao}, {Licata}, {Lindstr{\o}m}, {Lister}, {Livanou}, {Lobel},
  {L{\"o}ffler}, {L{\'o}pez}, {Lopez-Lozano}, {Lorenz}, {Loureiro},
  {MacDonald}, {Magalh{\~a}es Fernandes}, {Managau}, {Mann}, {Mantelet},
  {Marchal}, {Marchant}, {Marconi}, {Marie}, {Marinoni}, {Marrese},
  {Marschalk{\'o}}, {Marshall}, {Mart{\'\i}n-Fleitas}, {Martino}, {Mary},
  {Matijevi{\v{c}}}, {Mazeh}, {McMillan}, {Messina}, {Mestre}, {Michalik},
  {Millar}, {Miranda}, {Molina}, {Molinaro}, {Molinaro}, {Moln{\'a}r},
  {Moniez}, {Montegriffo}, {Monteiro}, {Mor}, {Mora}, {Morbidelli}, {Morel},
  {Morgenthaler}, {Morley}, {Morris}, {Mulone}, {Muraveva}, {Musella},
  {Narbonne}, {Nelemans}, {Nicastro}, {Noval}, {Ord{\'e}novic},
  {Ordieres-Mer{\'e}}, {Osborne}, {Pagani}, {Pagano}, {Pailler}, {Palacin},
  {Palaversa}, {Parsons}, {Paulsen}, {Pecoraro}, {Pedrosa}, {Pentik{\"a}inen},
  {Pereira}, {Pichon}, {Piersimoni}, {Pineau}, {Plachy}, {Plum}, {Poujoulet},
  {Pr{\v{s}}a}, {Pulone}, {Ragaini}, {Rago}, {Rambaux}, {Ramos-Lerate},
  {Ranalli}, {Rauw}, {Read}, {Regibo}, {Renk}, {Reyl{\'e}}, {Ribeiro},
  {Rimoldini}, {Ripepi}, {Riva}, {Rixon}, {Roelens}, {Romero-G{\'o}mez},
  {Rowell}, {Royer}, {Rudolph}, {Ruiz-Dern}, {Sadowski}, {Sagrist{\`a}
  Sell{\'e}s}, {Sahlmann}, {Salgado}, {Salguero}, {Sarasso}, {Savietto},
  {Schnorhk}, {Schultheis}, {Sciacca}, {Segol}, {Segovia}, {Segransan},
  {Serpell}, {Shih}, {Smareglia}, {Smart}, {Smith}, {Solano}, {Solitro},
  {Sordo}, {Soria Nieto}, {Souchay}, {Spagna}, {Spoto}, {Stampa}, {Steele},
  {Steidelm{\"u}ller}, {Stephenson}, {Stoev}, {Suess}, {S{\"u}veges}, {Surdej},
  {Szabados}, {Szegedi-Elek}, {Tapiador}, {Taris}, {Tauran}, {Taylor},
  {Teixeira}, {Terrett}, {Tingley}, {Trager}, {Turon}, {Ulla}, {Utrilla},
  {Valentini}, {van Elteren}, {Van Hemelryck}, {van Leeuwen}, {Varadi},
  {Vecchiato}, {Veljanoski}, {Via}, {Vicente}, {Vogt}, {Voss}, {Votruba},
  {Voutsinas}, {Walmsley}, {Weiler}, {Weingrill}, {Werner}, {Wevers},
  {Whitehead}, {Wyrzykowski}, {Yoldas}, {{\v{Z}}erjal}, {Zucker}, {Zurbach},
  {Zwitter}, {Alecu}, {Allen}, {Allende Prieto}, {Amorim},
  {Anglada-Escud{\'e}}, {Arsenijevic}, {Azaz}, {Balm}, {Beck}, {Bernstein},
  {Bigot}, {Bijaoui}, {Blasco}, {Bonfigli}, {Bono}, {Boudreault}, {Bressan},
  {Brown}, {Brunet}, {Bunclark}, {Buonanno}, {Butkevich}, {Carret}, {Carrion},
  {Chemin}, {Ch{\'e}reau}, {Corcione}, {Darmigny}, {de Boer}, {de Teodoro}, {de
  Zeeuw}, {Delle Luche}, {Domingues}, {Dubath}, {Fodor}, {Fr{\'e}zouls},
  {Fries}, {Fustes}, {Fyfe}, {Gallardo}, {Gallegos}, {Gardiol}, {Gebran},
  {Gomboc}, {G{\'o}mez}, {Grux}, {Gueguen}, {Heyrovsky}, {Hoar}, {Iannicola},
  {Isasi Parache}, {Janotto}, {Joliet}, {Jonckheere}, {Keil}, {Kim},
  {Klagyivik}, {Klar}, {Knude}, {Kochukhov}, {Kolka}, {Kos}, {Kutka}, {Lainey},
  {LeBouquin}, {Liu}, {Loreggia}, {Makarov}, {Marseille}, {Martayan},
  {Martinez-Rubi}, {Massart}, {Meynadier}, {Mignot}, {Munari}, {Nguyen},
  {Nordlander}, {Ocvirk}, {O'Flaherty}, {Olias Sanz}, {Ortiz}, {Osorio},
  {Oszkiewicz}, {Ouzounis}, {Palmer}, {Park}, {Pasquato}, {Peltzer}, {Peralta},
  {P{\'e}turaud}, {Pieniluoma}, {Pigozzi}, {Poels}, {Prat}, {Prod'homme},
  {Raison}, {Rebordao}, {Risquez}, {Rocca-Volmerange}, {Rosen}, {Ruiz-Fuertes},
  {Russo}, {Sembay}, {Serraller Vizcaino}, {Short}, {Siebert}, {Silva},
  {Sinachopoulos}, {Slezak}, {Soffel}, {Sosnowska}, {Strai{\v{z}}ys}, {ter
  Linden}, {Terrell}, {Theil}, {Tiede}, {Troisi}, {Tsalmantza}, {Tur},
  {Vaccari}, {Vachier}, {Valles}, {Van Hamme}, {Veltz}, {Virtanen}, {Wallut},
  {Wichmann}, {Wilkinson}, {Ziaeepour}, \& {Zschocke}}]{gaia2016}
{Gaia Collaboration}, {Prusti}, T., {de Bruijne}, J.~H.~J., {et~al.} 2016,
  \aap, 595, A1, \dodoi{10.1051/0004-6361/201629272}

\bibitem[{{Gaia Collaboration} {et~al.}(2018){Gaia Collaboration}, {Brown},
  {Vallenari}, {Prusti}, {de Bruijne}, {Babusiaux}, {Bailer-Jones}, {Biermann},
  {Evans}, {Eyer}, {Jansen}, {Jordi}, {Klioner}, {Lammers}, {Lindegren},
  {Luri}, {Mignard}, {Panem}, {Pourbaix}, {Randich}, {Sartoretti}, {Siddiqui},
  {Soubiran}, {van Leeuwen}, {Walton}, {Arenou}, {Bastian}, {Cropper},
  {Drimmel}, {Katz}, {Lattanzi}, {Bakker}, {Cacciari}, {Casta{\~n}eda},
  {Chaoul}, {Cheek}, {De Angeli}, {Fabricius}, {Guerra}, {Holl}, {Masana},
  {Messineo}, {Mowlavi}, {Nienartowicz}, {Panuzzo}, {Portell}, {Riello},
  {Seabroke}, {Tanga}, {Th{\'e}venin}, {Gracia-Abril}, {Comoretto},
  {Garcia-Reinaldos}, {Teyssier}, {Altmann}, {Andrae}, {Audard},
  {Bellas-Velidis}, {Benson}, {Berthier}, {Blomme}, {Burgess}, {Busso},
  {Carry}, {Cellino}, {Clementini}, {Clotet}, {Creevey}, {Davidson}, {De
  Ridder}, {Delchambre}, {Dell'Oro}, {Ducourant},
  {Fern{\'a}ndez-Hern{\'a}ndez}, {Fouesneau}, {Fr{\'e}mat}, {Galluccio},
  {Garc{\'\i}a-Torres}, {Gonz{\'a}lez-N{\'u}{\~n}ez}, {Gonz{\'a}lez-Vidal},
  {Gosset}, {Guy}, {Halbwachs}, {Hambly}, {Harrison}, {Hern{\'a}ndez},
  {Hestroffer}, {Hodgkin}, {Hutton}, {Jasniewicz}, {Jean-Antoine-Piccolo},
  {Jordan}, {Korn}, {Krone-Martins}, {Lanzafame}, {Lebzelter}, {L{\"o}ffler},
  {Manteiga}, {Marrese}, {Mart{\'\i}n-Fleitas}, {Moitinho}, {Mora}, {Muinonen},
  {Osinde}, {Pancino}, {Pauwels}, {Petit}, {Recio-Blanco}, {Richards},
  {Rimoldini}, {Robin}, {Sarro}, {Siopis}, {Smith}, {Sozzetti}, {S{\"u}veges},
  {Torra}, {van Reeven}, {Abbas}, {Abreu Aramburu}, {Accart}, {Aerts},
  {Altavilla}, {{\'A}lvarez}, {Alvarez}, {Alves}, {Anderson}, {Andrei},
  {Anglada Varela}, {Antiche}, {Antoja}, {Arcay}, {Astraatmadja}, {Bach},
  {Baker}, {Balaguer-N{\'u}{\~n}ez}, {Balm}, {Barache}, {Barata}, {Barbato},
  {Barblan}, {Barklem}, {Barrado}, {Barros}, {Barstow}, {Bartholom{\'e}
  Mu{\~n}oz}, {Bassilana}, {Becciani}, {Bellazzini}, {Berihuete}, {Bertone},
  {Bianchi}, {Bienaym{\'e}}, {Blanco-Cuaresma}, {Boch}, {Boeche}, {Bombrun},
  {Borrachero}, {Bossini}, {Bouquillon}, {Bourda}, {Bragaglia}, {Bramante},
  {Breddels}, {Bressan}, {Brouillet}, {Br{\"u}semeister}, {Brugaletta},
  {Bucciarelli}, {Burlacu}, {Busonero}, {Butkevich}, {Buzzi}, {Caffau},
  {Cancelliere}, {Cannizzaro}, {Cantat-Gaudin}, {Carballo}, {Carlucci},
  {Carrasco}, {Casamiquela}, {Castellani}, {Castro-Ginard}, {Charlot},
  {Chemin}, {Chiavassa}, {Cocozza}, {Costigan}, {Cowell}, {Crifo}, {Crosta},
  {Crowley}, {Cuypers}, {Dafonte}, {Damerdji}, {Dapergolas}, {David}, {David},
  {de Laverny}, {De Luise}, {De March}, {de Martino}, {de Souza}, {de Torres},
  {Debosscher}, {del Pozo}, {Delbo}, {Delgado}, {Delgado}, {Di Matteo},
  {Diakite}, {Diener}, {Distefano}, {Dolding}, {Drazinos}, {Dur{\'a}n},
  {Edvardsson}, {Enke}, {Eriksson}, {Esquej}, {Eynard Bontemps}, {Fabre},
  {Fabrizio}, {Faigler}, {Falc{\~a}o}, {Farr{\`a}s Casas}, {Federici},
  {Fedorets}, {Fernique}, {Figueras}, {Filippi}, {Findeisen}, {Fonti},
  {Fraile}, {Fraser}, {Fr{\'e}zouls}, {Gai}, {Galleti}, {Garabato},
  {Garc{\'\i}a-Sedano}, {Garofalo}, {Garralda}, {Gavel}, {Gavras}, {Gerssen},
  {Geyer}, {Giacobbe}, {Gilmore}, {Girona}, {Giuffrida}, {Glass}, {Gomes},
  {Granvik}, {Gueguen}, {Guerrier}, {Guiraud}, {Guti{\'e}rrez-S{\'a}nchez},
  {Haigron}, {Hatzidimitriou}, {Hauser}, {Haywood}, {Heiter}, {Helmi}, {Heu},
  {Hilger}, {Hobbs}, {Hofmann}, {Holland}, {Huckle}, {Hypki}, {Icardi},
  {Jan{\ss}en}, {Jevardat de Fombelle}, {Jonker}, {Juh{\'a}sz}, {Julbe},
  {Karampelas}, {Kewley}, {Klar}, {Kochoska}, {Kohley}, {Kolenberg},
  {Kontizas}, {Kontizas}, {Koposov}, {Kordopatis}, {Kostrzewa-Rutkowska},
  {Koubsky}, {Lambert}, {Lanza}, {Lasne}, {Lavigne}, {Le Fustec}, {Le
  Poncin-Lafitte}, {Lebreton}, {Leccia}, {Leclerc}, {Lecoeur-Taibi},
  {Lenhardt}, {Leroux}, {Liao}, {Licata}, {Lindstr{\o}m}, {Lister}, {Livanou},
  {Lobel}, {L{\'o}pez}, {Managau}, {Mann}, {Mantelet}, {Marchal}, {Marchant},
  {Marconi}, {Marinoni}, {Marschalk{\'o}}, {Marshall}, {Martino}, {Marton},
  {Mary}, {Massari}, {Matijevi{\v{c}}}, {Mazeh}, {McMillan}, {Messina},
  {Michalik}, {Millar}, {Molina}, {Molinaro}, {Moln{\'a}r}, {Montegriffo},
  {Mor}, {Morbidelli}, {Morel}, {Morris}, {Mulone}, {Muraveva}, {Musella},
  {Nelemans}, {Nicastro}, {Noval}, {O'Mullane}, {Ord{\'e}novic},
  {Ord{\'o}{\~n}ez-Blanco}, {Osborne}, {Pagani}, {Pagano}, {Pailler},
  {Palacin}, {Palaversa}, {Panahi}, {Pawlak}, {Piersimoni}, {Pineau}, {Plachy},
  {Plum}, {Poggio}, {Poujoulet}, {Pr{\v{s}}a}, {Pulone}, {Racero}, {Ragaini},
  {Rambaux}, {Ramos-Lerate}, {Regibo}, {Reyl{\'e}}, {Riclet}, {Ripepi}, {Riva},
  {Rivard}, {Rixon}, {Roegiers}, {Roelens}, {Romero-G{\'o}mez}, {Rowell},
  {Royer}, {Ruiz-Dern}, {Sadowski}, {Sagrist{\`a} Sell{\'e}s}, {Sahlmann},
  {Salgado}, {Salguero}, {Sanna}, {Santana-Ros}, {Sarasso}, {Savietto},
  {Schultheis}, {Sciacca}, {Segol}, {Segovia}, {S{\'e}gransan}, {Shih},
  {Siltala}, {Silva}, {Smart}, {Smith}, {Solano}, {Solitro}, {Sordo}, {Soria
  Nieto}, {Souchay}, {Spagna}, {Spoto}, {Stampa}, {Steele},
  {Steidelm{\"u}ller}, {Stephenson}, {Stoev}, {Suess}, {Surdej}, {Szabados},
  {Szegedi-Elek}, {Tapiador}, {Taris}, {Tauran}, {Taylor}, {Teixeira},
  {Terrett}, {Teyssandier}, {Thuillot}, {Titarenko}, {Torra Clotet}, {Turon},
  {Ulla}, {Utrilla}, {Uzzi}, {Vaillant}, {Valentini}, {Valette}, {van Elteren},
  {Van Hemelryck}, {van Leeuwen}, {Vaschetto}, {Vecchiato}, {Veljanoski},
  {Viala}, {Vicente}, {Vogt}, {von Essen}, {Voss}, {Votruba}, {Voutsinas},
  {Walmsley}, {Weiler}, {Wertz}, {Wevers}, {Wyrzykowski}, {Yoldas},
  {{\v{Z}}erjal}, {Ziaeepour}, {Zorec}, {Zschocke}, {Zucker}, {Zurbach}, \&
  {Zwitter}}]{gaia2018}
{Gaia Collaboration}, {Brown}, A.~G.~A., {Vallenari}, A., {et~al.} 2018, \aap,
  616, A1, \dodoi{10.1051/0004-6361/201833051}

\bibitem[{{Gallet}(2020)}]{florian2020}
{Gallet}, F. 2020, \aap, 641, A38, \dodoi{10.1051/0004-6361/202038058}

\bibitem[{{Gillon} {et~al.}(2012){Gillon}, {Triaud}, {Fortney}, {Demory},
  {Jehin}, {Lendl}, {Magain}, {Kabath}, {Queloz}, {Alonso}, {Anderson},
  {Collier Cameron}, {Fumel}, {Hebb}, {Hellier}, {Lanotte}, {Maxted},
  {Mowlavi}, \& {Smalley}}]{Gillon2012}
{Gillon}, M., {Triaud}, A.~H.~M.~J., {Fortney}, J.~J., {et~al.} 2012, \aap,
  542, A4, \dodoi{10.1051/0004-6361/201218817}

\bibitem[{{Gillon} {et~al.}(2014){Gillon}, {Anderson}, {Collier-Cameron},
  {Delrez}, {Hellier}, {Jehin}, {Lendl}, {Maxted}, {Pepe}, {Pollacco},
  {Queloz}, {S{\'e}gransan}, {Smith}, {Smalley}, {Southworth}, {Triaud},
  {Udry}, {Van Grootel}, \& {West}}]{Gillon2014}
{Gillon}, M., {Anderson}, D.~R., {Collier-Cameron}, A., {et~al.} 2014, \aap,
  562, L3, \dodoi{10.1051/0004-6361/201323014}

\bibitem[{{Goldreich} \& {Soter}(1966)}]{Goldreich1996}
{Goldreich}, P., \& {Soter}, S. 1966, \icarus, 5, 375,
  \dodoi{10.1016/0019-1035(66)90051-0}

\bibitem[{{Hartman} {et~al.}(2016){Hartman}, {Bakos}, {Bhatti}, {Penev},
  {Bieryla}, {Latham}, {Kov{\'a}cs}, {Torres}, {Csubry}, {de Val-Borro},
  {Buchhave}, {Kov{\'a}cs}, {Quinn}, {Howard}, {Isaacson}, {Fulton}, {Everett},
  {Esquerdo}, {B{\'e}ky}, {Szklenar}, {Falco}, {Santerne}, {Boisse},
  {H{\'e}brard}, {Burrows}, {L{\'a}z{\'a}r}, {Papp}, \&
  {S{\'a}ri}}]{Hartman2016}
{Hartman}, J.~D., {Bakos}, G.~{\'A}., {Bhatti}, W., {et~al.} 2016, \aj, 152,
  182, \dodoi{10.3847/0004-6256/152/6/182}

\bibitem[{{Hebb} {et~al.}(2009){Hebb}, {Collier-Cameron}, {Loeillet},
  {Pollacco}, {H{\'e}brard}, {Street}, {Bouchy}, {Stempels}, {Moutou},
  {Simpson}, {Udry}, {Joshi}, {West}, {Skillen}, {Wilson}, {McDonald},
  {Gibson}, {Aigrain}, {Anderson}, {Benn}, {Christian}, {Enoch}, {Haswell},
  {Hellier}, {Horne}, {Irwin}, {Lister}, {Maxted}, {Mayor}, {Norton}, {Parley},
  {Pont}, {Queloz}, {Smalley}, \& {Wheatley}}]{Hebb2009}
{Hebb}, L., {Collier-Cameron}, A., {Loeillet}, B., {et~al.} 2009, \apj, 693,
  1920, \dodoi{10.1088/0004-637X/693/2/1920}

\bibitem[{{Hebb} {et~al.}(2010){Hebb}, {Collier-Cameron}, {Triaud}, {Lister},
  {Smalley}, {Maxted}, {Hellier}, {Anderson}, {Pollacco}, {Gillon}, {Queloz},
  {West}, {Bentley}, {Enoch}, {Haswell}, {Horne}, {Mayor}, {Pepe}, {Segransan},
  {Skillen}, {Udry}, \& {Wheatley}}]{Hebb2010}
{Hebb}, L., {Collier-Cameron}, A., {Triaud}, A.~H.~M.~J., {et~al.} 2010, \apj,
  708, 224, \dodoi{10.1088/0004-637X/708/1/224}

\bibitem[{{Hellier} {et~al.}(2009){Hellier}, {Anderson}, {Collier Cameron},
  {Gillon}, {Hebb}, {Maxted}, {Queloz}, {Smalley}, {Triaud}, {West}, {Wilson},
  {Bentley}, {Enoch}, {Horne}, {Irwin}, {Lister}, {Mayor}, {Parley}, {Pepe},
  {Pollacco}, {Segransan}, {Udry}, \& {Wheatley}}]{Hellier2009}
{Hellier}, C., {Anderson}, D.~R., {Collier Cameron}, A., {et~al.} 2009, \nat,
  460, 1098, \dodoi{10.1038/nature08245}

\bibitem[{{Hellier} {et~al.}(2011){Hellier}, {Anderson}, {Collier Cameron},
  {Gillon}, {Jehin}, {Lendl}, {Maxted}, {Pepe}, {Pollacco}, {Queloz},
  {S{\'e}gransan}, {Smalley}, {Smith}, {Southworth}, {Triaud}, {Udry}, \&
  {West}}]{Hellier2011}
---. 2011, \aap, 535, L7, \dodoi{10.1051/0004-6361/201117081}

\bibitem[{{Henden} {et~al.}(2016){Henden}, {Templeton}, {Terrell}, {Smith},
  {Levine}, \& {Welch}}]{henden2016}
{Henden}, A.~A., {Templeton}, M., {Terrell}, D., {et~al.} 2016, VizieR Online
  Data Catalog, II/336

\bibitem[{{Hoyer} {et~al.}(2016){Hoyer}, {Pall{\'e}}, {Dragomir}, \&
  {Murgas}}]{Hoyer2016}
{Hoyer}, S., {Pall{\'e}}, E., {Dragomir}, D., \& {Murgas}, F. 2016, \aj, 151,
  137, \dodoi{10.3847/0004-6256/151/6/137}

\bibitem[{{Jenkins} {et~al.}(2016){Jenkins}, {Twicken}, {McCauliff},
  {Campbell}, {Sanderfer}, {Lung}, {Mansouri-Samani}, {Girouard}, {Tenenbaum},
  {Klaus}, {Smith}, {Caldwell}, {Chacon}, {Henze}, {Heiges}, {Latham},
  {Morgan}, {Swade}, {Rinehart}, \& {Vanderspek}}]{Jenkins2016}
{Jenkins}, J.~M., {Twicken}, J.~D., {McCauliff}, S., {et~al.} 2016, in Society
  of Photo-Optical Instrumentation Engineers (SPIE) Conference Series, Vol.
  9913, Software and Cyberinfrastructure for Astronomy IV, ed. G.~{Chiozzi} \&
  J.~C. {Guzman}, 99133E, \dodoi{10.1117/12.2233418}

\bibitem[{{Jiang} {et~al.}(2016){Jiang}, {Lai}, {Savushkin}, {Mkrtichian},
  {Antonyuk}, {Griv}, {Hsieh}, \& {Yeh}}]{Jiang2016}
{Jiang}, I.-G., {Lai}, C.-Y., {Savushkin}, A., {et~al.} 2016, \aj, 151, 17,
  \dodoi{10.3847/0004-6256/151/1/17}

\bibitem[{{Kempton} {et~al.}(2018){Kempton}, {Bean}, {Louie}, {Deming}, {Koll},
  {Mansfield}, {Christiansen}, {L{\'o}pez-Morales}, {Swain}, {Zellem},
  {Ballard}, {Barclay}, {Barstow}, {Batalha}, {Beatty}, {Berta-Thompson},
  {Birkby}, {Buchhave}, {Charbonneau}, {Cowan}, {Crossfield}, {de Val-Borro},
  {Doyon}, {Dragomir}, {Gaidos}, {Heng}, {Hu}, {Kane}, {Kreidberg}, {Mallonn},
  {Morley}, {Narita}, {Nascimbeni}, {Pall{\'e}}, {Quintana}, {Rauscher},
  {Seager}, {Shkolnik}, {Sing}, {Sozzetti}, {Stassun}, {Valenti}, \& {von
  Essen}}]{Kempton2018}
{Kempton}, E. M.~R., {Bean}, J.~L., {Louie}, D.~R., {et~al.} 2018, \pasp, 130,
  114401, \dodoi{10.1088/1538-3873/aadf6f}

\bibitem[{{Kov{\'a}cs} {et~al.}(2002){Kov{\'a}cs}, {Zucker}, \&
  {Mazeh}}]{Kovacs2002}
{Kov{\'a}cs}, G., {Zucker}, S., \& {Mazeh}, T. 2002, \aap, 391, 369,
  \dodoi{10.1051/0004-6361:20020802}

\bibitem[{{Lenz} \& {Breger}(2005)}]{Lenz2005}
{Lenz}, P., \& {Breger}, M. 2005, Communications in Asteroseismology, 146, 53,
  \dodoi{10.1553/cia146s53}

\bibitem[{{Levrard} {et~al.}(2009){Levrard}, {Winisdoerffer}, \&
  {Chabrier}}]{Levrard2009}
{Levrard}, B., {Winisdoerffer}, C., \& {Chabrier}, G. 2009, \apjl, 692, L9,
  \dodoi{10.1088/0004-637X/692/1/L9}

\bibitem[{{Lightkurve Collaboration} {et~al.}(2018){Lightkurve Collaboration},
  {Cardoso}, {Hedges}, {Gully-Santiago}, {Saunders}, {Cody}, {Barclay}, {Hall},
  {Sagear}, {Turtelboom}, {Zhang}, {Tzanidakis}, {Mighell}, {Coughlin}, {Bell},
  {Berta-Thompson}, {Williams}, {Dotson}, \& {Barentsen}}]{Cardoso2018}
{Lightkurve Collaboration}, {Cardoso}, J. V. d.~M., {Hedges}, C., {et~al.}
  2018, {Lightkurve: Kepler and TESS time series analysis in Python}.
\newblock \doeprint{1812.013}

\bibitem[{Lomb(1976)}]{lomb1976}
Lomb, N.~R. 1976, Astrophysics and space science, 39, 447

\bibitem[{{Maciejewski} {et~al.}(2013){Maciejewski}, {Puchalski}, {Saral},
  {Derman}, {Kitze}, {Bukowiecki}, {Seeliger}, \&
  {Neuhaeuser}}]{Maciejewski2013}
{Maciejewski}, G., {Puchalski}, D., {Saral}, G., {et~al.} 2013, Information
  Bulletin on Variable Stars, 6082, 1

\bibitem[{{Martins} {et~al.}(2020){Martins}, {Gomes}, {Messias}, {de Lira},
  {Le{\~a}o}, {Almeida}, {Teixeira}, {das Chagas}, {Bravo}, {Belete}, \& {De
  Medeiros}}]{Martins2020}
{Martins}, B.~L.~C., {Gomes}, R.~L., {Messias}, Y.~S., {et~al.} 2020, \apjs,
  250, 20, \dodoi{10.3847/1538-4365/aba73f}

\bibitem[{{Molli{\`e}re} {et~al.}(2017){Molli{\`e}re}, {van Boekel}, {Bouwman},
  {Henning}, {Lagage}, \& {Min}}]{Moliere2017}
{Molli{\`e}re}, P., {van Boekel}, R., {Bouwman}, J., {et~al.} 2017, \aap, 600,
  A10, \dodoi{10.1051/0004-6361/201629800}

\bibitem[{{Murgas} {et~al.}(2014){Murgas}, {Pall{\'e}}, {Zapatero Osorio},
  {Nortmann}, {Hoyer}, \& {Cabrera-Lavers}}]{Murgas2014}
{Murgas}, F., {Pall{\'e}}, E., {Zapatero Osorio}, M.~R., {et~al.} 2014, \aap,
  563, A41, \dodoi{10.1051/0004-6361/201322374}

\bibitem[{{Newville} {et~al.}(2016){Newville}, {Stensitzki}, {Allen}, {Rawlik},
  {Ingargiola}, \& {Nelson}}]{Newville2016}
{Newville}, M., {Stensitzki}, T., {Allen}, D.~B., {et~al.} 2016, {Lmfit:
  Non-Linear Least-Square Minimization and Curve-Fitting for Python}.
\newblock \doeprint{1606.014}

\bibitem[{{Ogilvie}(2014)}]{Ogilvie2014}
{Ogilvie}, G.~I. 2014, \araa, 52, 171,
  \dodoi{10.1146/annurev-astro-081913-035941}

\bibitem[{{Patra} {et~al.}(2017){Patra}, {Winn}, {Holman}, {Yu}, {Deming}, \&
  {Dai}}]{patra2017}
{Patra}, K.~C., {Winn}, J.~N., {Holman}, M.~J., {et~al.} 2017, \aj, 154, 4,
  \dodoi{10.3847/1538-3881/aa6d75}

\bibitem[{{Patra} {et~al.}(2020){Patra}, {Winn}, {Holman}, {Gillon},
  {Burdanov}, {Jehin}, {Delrez}, {Pozuelos}, {Barkaoui}, {Benkhaldoun},
  {Narita}, {Fukui}, {Kusakabe}, {Kawauchi}, {Terada}, {Bouma}, {Weinberg}, \&
  {Broome}}]{Patra2020}
---. 2020, \aj, 159, 150, \dodoi{10.3847/1538-3881/ab7374}

\bibitem[{{Penev} {et~al.}(2018){Penev}, {Bouma}, {Winn}, \&
  {Hartman}}]{Penev2018}
{Penev}, K., {Bouma}, L.~G., {Winn}, J.~N., \& {Hartman}, J.~D. 2018, \aj, 155,
  165, \dodoi{10.3847/1538-3881/aaaf71}

\bibitem[{{Plavchan} {et~al.}(2008){Plavchan}, {Jura}, {Kirkpatrick}, {Cutri},
  \& {Gallagher}}]{Plavchan2008}
{Plavchan}, P., {Jura}, M., {Kirkpatrick}, J.~D., {Cutri}, R.~M., \&
  {Gallagher}, S.~C. 2008, \apjs, 175, 191, \dodoi{10.1086/523644}

\bibitem[{{Ricci} {et~al.}(2015){Ricci}, {Ram{\'o}n-Fox}, {Ayala-Loera},
  {Michel}, {Navarro-Meza}, {Fox-Machado}, {Reyes-Ruiz}, {Brown Sevilla}, \&
  {Curiel}}]{Ricci2015}
{Ricci}, D., {Ram{\'o}n-Fox}, F.~G., {Ayala-Loera}, C., {et~al.} 2015, \pasp,
  127, 143, \dodoi{10.1086/680233}

\bibitem[{{Salvatier} {et~al.}(2016){Salvatier}, {Wiecki{\^a}}, \&
  {Fonnesbeck}}]{Salvatier2016}
{Salvatier}, J., {Wiecki{\^a}}, T.~V., \& {Fonnesbeck}, C. 2016, {PyMC3: Python
  probabilistic programming framework}.
\newblock \doeprint{1610.016}

\bibitem[{{Salz} {et~al.}(2015){Salz}, {Schneider}, {Czesla}, \&
  {Schmitt}}]{Salz2015}
{Salz}, M., {Schneider}, P.~C., {Czesla}, S., \& {Schmitt}, J.~H.~M.~M. 2015,
  \aap, 576, A42, \dodoi{10.1051/0004-6361/201425243}

\bibitem[{{Sanz-Forcada} {et~al.}(2011){Sanz-Forcada}, {Micela}, {Ribas},
  {Pollock}, {Eiroa}, {Velasco}, {Solano}, \&
  {Garc{\'\i}a-{\'A}lvarez}}]{Sanz2011}
{Sanz-Forcada}, J., {Micela}, G., {Ribas}, I., {et~al.} 2011, \aap, 532, A6,
  \dodoi{10.1051/0004-6361/201116594}

\bibitem[{{Scargle}(1982)}]{Scargle1982}
{Scargle}, J.~D. 1982, \apj, 263, 835, \dodoi{10.1086/160554}

\bibitem[{{Schlegel} {et~al.}(1998){Schlegel}, {Finkbeiner}, \&
  {Davis}}]{schlegel1998}
{Schlegel}, D.~J., {Finkbeiner}, D.~P., \& {Davis}, M. 1998, \apj, 500, 525,
  \dodoi{10.1086/305772}

\bibitem[{{Sousa} {et~al.}(2018){Sousa}, {Adibekyan}, {Delgado-Mena}, {Santos},
  {Andreasen}, {Ferreira}, {Tsantaki}, {Barros}, {Demangeon}, {Israelian},
  {Faria}, {Figueira}, {Mortier}, {Brand{\~a}o}, {Montalto}, {Rojas-Ayala}, \&
  {Santerne}}]{sousa2018}
{Sousa}, S.~G., {Adibekyan}, V., {Delgado-Mena}, E., {et~al.} 2018, \aap, 620,
  A58, \dodoi{10.1051/0004-6361/201833350}

\bibitem[{{Southworth} {et~al.}(2019){Southworth}, {Dominik}, {J{\o}rgensen},
  {Andersen}, {Bozza}, {Burgdorf}, {D'Ago}, {Dib}, {Figuera Jaimes}, {Fujii},
  {Gill}, {Haikala}, {Hinse}, {Hundertmark}, {Khalouei}, {Korhonen},
  {Longa-Pe{\~n}a}, {Mancini}, {Peixinho}, {Rabus}, {Rahvar}, {Sajadian},
  {Skottfelt}, {Snodgrass}, {Spyratos}, {Tregloan-Reed}, {Unda-Sanzana}, \&
  {von Essen}}]{Southworth2019}
{Southworth}, J., {Dominik}, M., {J{\o}rgensen}, U.~G., {et~al.} 2019, \mnras,
  490, 4230, \dodoi{10.1093/mnras/stz2602}

\bibitem[{{Stassun} \& {Torres}(2018)}]{stas2018}
{Stassun}, K.~G., \& {Torres}, G. 2018, \apj, 862, 61,
  \dodoi{10.3847/1538-4357/aacafc}

\bibitem[{{Stevenson} {et~al.}(2014){Stevenson}, {D{\'e}sert}, {Line}, {Bean},
  {Fortney}, {Showman}, {Kataria}, {Kreidberg}, {McCullough}, {Henry},
  {Charbonneau}, {Burrows}, {Seager}, {Madhusudhan}, {Williamson}, \&
  {Homeier}}]{Stevenson2014}
{Stevenson}, K.~B., {D{\'e}sert}, J.-M., {Line}, M.~R., {et~al.} 2014, Science,
  346, 838, \dodoi{10.1126/science.1256758}

\bibitem[{{Stevenson} {et~al.}(2017){Stevenson}, {Line}, {Bean}, {D{\'e}sert},
  {Fortney}, {Showman}, {Kataria}, {Kreidberg}, \& {Feng}}]{Stevenson2017}
{Stevenson}, K.~B., {Line}, M.~R., {Bean}, J.~L., {et~al.} 2017, \aj, 153, 68,
  \dodoi{10.3847/1538-3881/153/2/68}

\bibitem[{{Su{\'a}rez Mascare{\~n}o} {et~al.}(2015){Su{\'a}rez Mascare{\~n}o},
  {Rebolo}, {Gonz{\'a}lez Hern{\'a}ndez}, \& {Esposito}}]{suarez2015}
{Su{\'a}rez Mascare{\~n}o}, A., {Rebolo}, R., {Gonz{\'a}lez Hern{\'a}ndez},
  J.~I., \& {Esposito}, M. 2015, \mnras, 452, 2745,
  \dodoi{10.1093/mnras/stv1441}

\bibitem[{{Sun} {et~al.}(2018){Sun}, {Ji}, \& {Dong}}]{Zhao2018}
{Sun}, Z., {Ji}, J.-h., \& {Dong}, Y. 2018, \caa, 42, 101,
  \dodoi{10.1016/j.chinastron.2018.01.007}

\bibitem[{{Tayar} {et~al.}(2020){Tayar}, {Claytor}, {Huber}, \& {van
  Saders}}]{tayar2020}
{Tayar}, J., {Claytor}, Z.~R., {Huber}, D., \& {van Saders}, J. 2020, arXiv
  e-prints, arXiv:2012.07957.
\newblock \doarXiv{2012.07957}

\bibitem[{{Tsantaki} {et~al.}(2013){Tsantaki}, {Sousa}, {Adibekyan}, {Santos},
  {Mortier}, \& {Israelian}}]{tsantaki2013}
{Tsantaki}, M., {Sousa}, S.~G., {Adibekyan}, V.~Z., {et~al.} 2013, \aap, 555,
  A150, \dodoi{10.1051/0004-6361/201321103}

\bibitem[{{VanderPlas} \& {Ivezi{\'c}}(2015)}]{VanderPlas2015}
{VanderPlas}, J.~T., \& {Ivezi{\'c}}, {\v{Z}}. 2015, \apj, 812, 18,
  \dodoi{10.1088/0004-637X/812/1/18}

\bibitem[{Virtanen {et~al.}(2020)Virtanen, Gommers, Oliphant, Haberland, Reddy,
  Cournapeau, Burovski, Peterson, Weckesser, Bright, {van der Walt}, Brett,
  Wilson, Millman, Mayorov, Nelson, Jones, Kern, Larson, Carey, Polat, Feng,
  Moore, {VanderPlas}, Laxalde, Perktold, Cimrman, Henriksen, Quintero, Harris,
  Archibald, Ribeiro, Pedregosa, {van Mulbregt}, \& {SciPy 1.0
  Contributors}}]{Virtanen2020}
Virtanen, P., Gommers, R., Oliphant, T.~E., {et~al.} 2020, Nature Methods, 17,
  261, \dodoi{10.1038/s41592-019-0686-2}

\bibitem[{Wang \& Dai(2020)}]{Lile2020}
Wang, L., \& Dai, F. 2020, arXiv preprint arXiv:2101.00042

\bibitem[{{Weaver} {et~al.}(2020){Weaver}, {L{\'o}pez-Morales}, {Espinoza},
  {Rackham}, {Osip}, {Apai}, {Jord{\'a}n}, {Bixel}, {Lewis}, {Alam}, {Kirk},
  {McGruder}, {Rodler}, \& {Fienco}}]{Weaver2020}
{Weaver}, I.~C., {L{\'o}pez-Morales}, M., {Espinoza}, N., {et~al.} 2020, \aj,
  159, 13, \dodoi{10.3847/1538-3881/ab55da}

\bibitem[{{Wilson} {et~al.}(2008){Wilson}, {Gillon}, {Hellier}, {Maxted},
  {Pepe}, {Queloz}, {Anderson}, {Collier Cameron}, {Smalley}, {Lister},
  {Bentley}, {Blecha}, {Christian}, {Enoch}, {Haswell}, {Hebb}, {Horne},
  {Irwin}, {Joshi}, {Kane}, {Marmier}, {Mayor}, {Parley}, {Pollacco}, {Pont},
  {Ryans}, {Segransan}, {Skillen}, {Street}, {Udry}, {West}, \&
  {Wheatley}}]{Wilson2008}
{Wilson}, D.~M., {Gillon}, M., {Hellier}, C., {et~al.} 2008, \apjl, 675, L113,
  \dodoi{10.1086/586735}

\bibitem[{{Winn} {et~al.}(2008){Winn}, {Holman}, {Torres}, {McCullough},
  {Johns-Krull}, {Latham}, {Shporer}, {Mazeh}, {Garcia-Melendo}, {Foote},
  {Esquerdo}, \& {Everett}}]{winn2008}
{Winn}, J.~N., {Holman}, M.~J., {Torres}, G., {et~al.} 2008, \apj, 683, 1076,
  \dodoi{10.1086/589737}

\bibitem[{{Wong} {et~al.}(2020){Wong}, {Shporer}, {Daylan}, {Benneke},
  {Fetherolf}, {Kane}, {Ricker}, {Vanderspek}, {Latham}, {Winn}, {Jenkins},
  {Boyd}, {Glidden}, {Goeke}, {Sha}, {Ting}, \& {Yahalomi}}]{Wong2020}
{Wong}, I., {Shporer}, A., {Daylan}, T., {et~al.} 2020, \aj, 160, 155,
  \dodoi{10.3847/1538-3881/ababad}

\bibitem[{{Yee} {et~al.}(2020){Yee}, {Winn}, {Knutson}, {Patra},
  {Vissapragada}, {Zhang}, {Holman}, {Shporer}, \& {Wright}}]{Yee2020}
{Yee}, S.~W., {Winn}, J.~N., {Knutson}, H.~A., {et~al.} 2020, \apjl, 888, L5,
  \dodoi{10.3847/2041-8213/ab5c16}

\end{thebibliography}

\clearpage
\appendix
\label{sec:appendix}

Appendix: 
Limb darkening parameters of global modeling (Table-\ref{tab:wave}) and the details of TTV analysis (Table-\ref{ttvanalysis}).

\setcounter{table}{0}
\renewcommand{\thetable}{A\arabic{table}}
\begin{deluxetable}{cccccc}[!hb]
\setlength{\tabcolsep}{5pt}
\renewcommand{\arraystretch}{0.8}
\tablecaption{Limb darkening parameters from the global modelling. \label{tab:wave}}
\tablehead{\colhead{~~~Parameter} & \colhead{Units} & \multicolumn{4}{c}{Values}}
\startdata
\tiny
\smallskip\\\multicolumn{2}{l}{Wavelength Parameters:}&J&H&K&$4.5\mu m$\smallskip\\
~~~~$u_{1}$\dotfill & linear limb-darkening coeff \dotfill &$0.152\pm0.045$&$0.091^{+0.040}_{-0.039}$&$0.088^{+0.040}_{-0.039}$&$0.055^{+0.037}_{-0.032}$\\
~~~~$u_{2}$\dotfill & quadratic limb-darkening coeff \dotfill &$0.231\pm0.045$&$0.318\pm0.044$&$0.284^{+0.043}_{-0.044}$&$0.183^{+0.041}_{-0.042}$\\
\smallskip\\\multicolumn{2}{l}{}&g'&r'&i'&z'\smallskip\\
~~~~$u_{1}$\dotfill & linear limb-darkening coeff \dotfill &$0.744\pm0.059$&$0.519^{+0.048}_{-0.049}$&$0.367^{+0.045}_{-0.046}$&$0.279\pm0.047$\\
~~~~$u_{2}$\dotfill & quadratic limb-darkening coeff \dotfill &$0.101^{+0.068}_{-0.069}$&$0.244^{+0.055}_{-0.054}$&$0.287\pm0.049$&$0.276^{+0.048}_{-0.049}$\\
\smallskip\\\multicolumn{2}{l}{}&V&R&I&TESS\smallskip\\
~~~~$u_{1}$\dotfill & linear limb-darkening coeff \dotfill &$0.601^{+0.063}_{-0.064}$&$0.540^{+0.052}_{-0.053}$&$0.354\pm0.048$&$0.426\pm0.037$\\
~~~~$u_{2}$\dotfill & quadratic limb-darkening coeff \dotfill &$0.162\pm0.064$&$0.266\pm0.056$&$0.284\pm0.050$&$0.275\pm0.040$\\
\enddata
\end{deluxetable}

\begin{deluxetable*}{ccl}[ht]
\tablewidth{0pt}
\tablecaption{Details of the TTV diagram and some transit parameters derived from {\sc exofast-v1}. \label{ttvanalysis}}
\tablehead{\colhead{Label} & \colhead{Unit} & \colhead{Description}}
\startdata
   MidTrTime        &  d    & Measured Mid-Transit Time  \\
   MidTrTimeErr     &  d    & Error of the Measured Mid-Transit Time  \\
   MidTrTimeOrig    &  d    & Observer-Provided Mid-Transit Time  \\
   MidTrTimeOrigErr &  d    & Error of the Observer-Provided Mid-Transit Time  \\
   Epoch            &  ---  & Epoch of Observation  \\
   O-C              &  d    & O-C  \\
   LineResiduals    &  d    & Residuals From The Linear Fit  \\
   Passband         &  ---  & Passband  \\
   LCType           &  ---  & Light Curve Type  \\
   Delta            &  ---  & Transit Depth  \\
 e\_Delta            &  ---  & Error of the Transit Depth  \\
   T14              &  d    & Transit Duration  \\
   FWHM             &  d    & FWHM Duration  \\
 e\_FWHM             &  d    & Error of the FWHM Duration  \\
   Tau              &  d    & Ingress/egress duration  \\
 e\_Tau              &  d    & Error of the Ingress/egress duration  \\
   Beta             &  ---  & Beta (Red Noise Parameter)  \\
   PNR              &  ---  & PNR (White Noise Parameter in ppt)  \\
   RMS              &  ---  & RMS Error from the Model  \\
   Discarded        &  ---  & Discarded  \\
   Outlier          &  ---  & Outlier not included in TTV analysis  \\
\enddata
\tablecomments{Table \ref{ttvanalysis} is published in its entirety in the 
machine-readable format. A portion is shown here for guidance regarding its form and content.}
\end{deluxetable*}

\end{document}